\documentclass[12pt]{article}
\usepackage{amsmath}
\usepackage{amssymb}
\usepackage{graphicx}
\usepackage{mathrsfs}
\font\blackboard=msbm10 at 12pt
\font\blackboards=msbm7
\font\blackboardss=msbm5
\newfam\black
\textfont\black=\blackboard
\scriptfont\black=\blackboards
\scriptscriptfont\black=\blackboardss

\newcommand{\junk}[1]{}

\newcommand{\ba}{\begin{array}}
\newcommand{\ea}{\end{array}}
\newcommand{\be}{\begin{equation}}
\newcommand{\ee}{\end{equation}}
\newcommand{\bea}{\begin{eqnarray}}
\newcommand{\eea}{\end{eqnarray}}
\newcommand{\beas}{\begin{eqnarray*}}
\newcommand{\eeas}{\end{eqnarray*}}

\def\laplace{{\kern1pt\vbox{\hrule height 1.2pt\hbox{\vrule width
1.2pt\hskip
  3pt\vbox{\vskip 6pt}\hskip 3pt\vrule width 0.6pt}\hrule height
  0.6pt}
  \kern1pt}}
\def\scriptlap{{\kern1pt\vbox{\hrule height 0.8pt\hbox{\vrule width
  0.8pt
  \hskip2pt\vbox{\vskip 4pt}\hskip 2pt\vrule width 0.4pt}\hrule height
  0.4pt}
  \kern1pt}}

\def\roughly#1{\raise.3ex\hbox{$#1$\kern-.75em\lower1ex\hbox{$\sim$}}}

\numberwithin{equation}{section}

\newcommand{\gone}[1]{}

\begin{document}
\pagestyle{plain}
\setcounter{page}{1}

\baselineskip16pt

\begin{titlepage}

\begin{flushright}

\end{flushright}
\vspace{8 mm}

\begin{center}

{\Large \bf Group manifold approach to higher spin theory  \\}

\end{center}

\vspace{7 mm}

\begin{center}

{\bf Shan Hu $^{a}$\footnote{E-mail:hushan@itp.ac.cn}, and Tianjun Li $^{a,b}$\footnote{E-mail:tli@itp.ac.cn}}

\vspace{3mm}
{\baselineskip 20pt \it $^a$
State Key Laboratory of Theoretical Physics and Kavli Institute for Theoretical Physics China (KITPC),
Institute of Theoretical Physics, Chinese Academy of Sciences, Beijing 100190, P. R. China \\
}
{\it $^b$
School of Physical Electronics, University of Electronic Science and Technology of China,\\
Chengdu 610054, P. R. China \\
}

\vspace{3mm}

\end{center}

\vspace{8 mm}

\begin{abstract}

We consider the group manifold approach to higher spin theory. The deformed local higher spin transformation is realized as the diffeomorphism transformation in the group manifold $\textbf{M}$. With the suitable rheonomy condition and the torsion constraint imposed, the unfolded equation can be obtained from the Bianchi identity, by solving which, fields in $\textbf{M}$ are determined by the multiplet at one point, or equivalently, by $(W^{[a(s-1),b(0)]}_{\mu},H)$ in $AdS_{4}\subset \textbf{M}$. Although the space is extended to $\textbf{M}$ to get the geometrical formulation, the dynamical degrees of freedom are still in $AdS_{4}$. The $4d$ equations of motion for $(W^{[a(s-1),b(0)]}_{\mu},H)$ are obtained by plugging the rheonomy condition into the Bianchi identity. The proper rheonomy condition allowing for the maximum on-shell degrees of freedom is given by Vasiliev equation. We also discuss the theory with the global higher spin symmetry, which is in parallel with the WZ model in supersymmetry.

\end{abstract}

\vspace{1cm}

\begin{flushleft}

\end{flushleft}
\end{titlepage}
\newpage

\section{Introduction}

Group manifold approach provides a natural geometrical formulation for supergravity \cite{1ggg,11gg,1g,2h}. The starting point is the supergroup $\overline{Osp(1/4)}$ or $Osp(1/4)$. Supergravity field and matter field are vielbein 1-form $\nu^{A}_{\bar{M}}$ and 0-form $H$ on the group manifold $\textbf{M}$, $A,\bar{M}=1,\cdots,\dim \overline{Osp(1/4)}$. Local super Poincare transformation is realized as the diffeomorphism transformation on $\textbf{M}$. The curvature $R^{A}_{\bar{M}\bar{N}}$ for the 1-form can be defined, on which, the rheonomy condition is imposed \cite{1ggg,11gg,1g,2h}. The condition requires that $R^{A}_{\bar{M}\bar{N}}$ can be algebraically expressed in terms of its purely ``inner'' components $R^{A}_{\mu\nu}$ with $\mu,\nu=1,2,3,4$ the indices in a four-dimensional submanifold $M_{4}$. Namely, 
\begin{equation}\label{11}
	R^{A}_{\bar{M}\bar{N}} = r^{A}_{\bar{M}\bar{N}}|^{\mu\nu}_{B}R^{B}_{\mu\nu},\;\;\;\;\;\;\;\;\;\textnormal{or}\;\;\;\;\;\;\;\;\;R^{A}_{CD} = r^{A}_{CD}|^{ab}_{B}R^{B}_{ab}
\end{equation}
where $r^{A}_{\bar{M}\bar{N}}|^{\mu\nu}_{B}$ and $r^{A}_{CD}|^{ab}_{B}$ are constant holonomic and anholonomic tensors. $a,b=1,2,3,4$. The rheonomy condition ensures that fields on the whole $\textbf{M}$ are determined by fields on $M_{4}$. So the final dynamics is still in $M_{4}$, where the diffeomorphism transformation reduces to the on-shell super Poincare transformation of the $4d$ fields. The equations of motion in $M_{4}$ are obtained by plugging the rheonomy condition into the Bianchi identity. Instead of imposing the rheonomy condition, one can also construct the extended action, which is the integration of some 4-form on a $4d$ submanifold $M_{4}$. Variation of the action with respect to both fields and $M_{4}$ gives the rheonomy condition as well as the $4d$ equations of motion.

In this paper, we will reformulate the group manifold method, adding an infinite number of auxiliary fields so that the final system is equivalent to the unfolded dynamics approach which is convenient for higher spin theory \cite{3h}. For simplicity, we will consider the minimal bosonic $4d$ HS algebra $ho(1|2:[3,2])$ with spin $s=0,2,\cdots$ \cite{mi}. The corresponding group manifold is denoted as $\textbf{M}$. Fields are 1-form $W_{\bar{M}}^{\alpha}$ and 0-form $H$ on $\textbf{M}$ with the curvature 2-form
\begin{equation}\label{1azd4}
	d W^{\alpha} = \frac{1}{2}(f^{\alpha}_{\beta\gamma}+R^{\alpha}_{\beta\gamma})W^{\beta} \wedge W^{\gamma}= \frac{1}{2}\hat{f}^{\alpha}_{\beta\gamma}W^{\beta} \wedge W^{\gamma}
\end{equation}
and the 1-form 
\begin{equation}\label{1azd44}
	d H = H_{\alpha}W^{\alpha}.
\end{equation}
$\bar{M} = 1,2,\cdots,\dim ho(1|2:[3,2])$. $\alpha \sim [a(s-1),b(t)]$ is in the adjoint representation of $ho(1|2:[3,2])$. $f^{\alpha}_{\beta\gamma}$ is the structure constant of $ho(1|2:[3,2])$. The deformed higher spin transformation is the diffeomorphism transformation on $\textbf{M}$.

The rheonomy condition is
\begin{eqnarray}\label{1azd5}
\nonumber && \hat{f}^{\alpha}_{\beta\gamma}=\hat{f}^{\alpha}_{\beta\gamma} (R^{[a(s-1),b(s-1)]}_{ab},R^{[a(s-1),b(s-1)]}_{ab;c_{1}},\cdots,H,H_{c_{1}},\cdots)
 \\ && H_{\alpha}=h_{\alpha} (R^{[a(s-1),b(s-1)]}_{ab},R^{[a(s-1),b(s-1)]}_{ab;c_{1}},\cdots,H,H_{c_{1}},\cdots),
\end{eqnarray}
where 
\begin{equation}\label{where}
	R^{[a(s-1),b(s-1)]}_{ab;c_{1}\cdots c_{n}} = \partial_{c_{n}}\cdots \partial_{c_{1}} R^{[a(s-1),b(s-1)]}_{ab},\;\;\;\;\;\;\;\;\;H_{c_{1}\cdots c_{n}} =\partial_{c_{n}}\cdots \partial_{c_{1}} H.
\end{equation}
$\partial_{c} = W^{\bar{M}}_{c} \partial_{\bar{M}}$, $a,b,c=1,2,3,4$. $a$ is the abbreviation for the $[0,a]$ element of $ho(1|2:[3,2])$. Different from the supergravity situation, the curvature depends on the ``inner'' components as well as their ``inner'' derivatives. This is the most generic rheonomy condition. (\ref{1azd5}) together with the Bianchi identity gives the unfolded equation 
\begin{eqnarray}
\nonumber & & d W^{\alpha} =\frac{1}{2}\hat{f}^{\alpha}_{\beta\gamma}W^{\beta}\wedge W^{\gamma},  \\ \nonumber & & d R^{[a(s-1),b(s-1)]}_{ab;c_{1}\cdots c_{n}} = r^{[a(s-1),b(s-1)]}_{ab;c_{1}\cdots c_{n} \gamma}W^{\gamma},\\  & & d H_{c_{1}\cdots c_{n}} = h_{c_{1}\cdots c_{n} \gamma}W^{\gamma},  
\end{eqnarray}
from which, $(W^{\alpha}, R^{[a(s-1),b(s-1)]}_{ab;c_{1}\cdots c_{n}},H_{c_{1}\cdots c_{n}})$ on the whole $\textbf{M}$ is determined by its value at one point. On $AdS_{4}\subset \textbf{M}$, we have the further relation 
\begin{eqnarray}\label{5ft78h}
  \nonumber && 	(R^{[a(s-1),b(s-1)]}_{ab},R^{[a(s-1),b(s-1)]}_{ab;c_{1}},\cdots,H,H_{c_{1}},\cdots)   \\ &\sim &(W^{[a(s-1),b(0)]}_{\mu},\partial_{\nu_{1}}W^{[a(s-1),b(0)]}_{\mu},\cdots,H,\partial_{\nu_{1}}H,\cdots),
\end{eqnarray}
where $\partial_{\mu}$ is the derivative on $AdS_{4}$. So equivalently, with $(W^{[a(s-1),b(0)]}_{\mu},H)$ given on $AdS_{4}$, $(W^{\alpha}_{\bar{M}},H)$ on the whole $\textbf{M}$ can be determined up to a gauge transformation. The dynamical 1-form fields are $W^{[a(s-1),b(0)]}$, which is because in (\ref{1azd5}), the torsion constraint is also implicitly imposed: $\hat{f}^{\alpha}_{\beta\gamma}$ and $H_{\alpha}$ do not depend on $R^{[a(s-1),b(t)]}_{ab;c_{1}\cdots c_{n}}$ with $t \neq s-1$. For 0-form, the deformed higher spin transformation is $\xi^{\bar{M}}\partial_{\bar{M}}=\epsilon^{\alpha}\partial_{\alpha}$, under which, the multiplet $(R^{[a(s-1),b(s-1)]}_{ab},R^{[a(s-1),b(s-1)]}_{ab;c_{1}},\cdots,H,H_{c_{1}},\cdots)$ forms the complete representation on-shell.

The whole dynamics is encoded in functions $(\hat{f}^{\alpha}_{\beta\gamma},h_{\alpha})$, which should satisfy the Bianchi identity and also give the correct free theory limit. With the unfolded equation plugged in, the Bianchi identities are polynomials of $(R^{[a(s-1),b(s-1)]}_{ab;c_{1}\cdots c_{n}},H_{c_{1}\cdots c_{n}})$, by solving which, $(\hat{f}^{\alpha}_{\beta\gamma},h_{\alpha})$ is determined with the rest constraints on $(R^{[a(s-1),b(s-1)]}_{ab;c_{1}\cdots c_{n}},H_{c_{1}\cdots c_{n}})$ acting as the $4d$ equations of motion. The procedure is simple in supergravity but is extremely complicated in higher spin theory. Instead of fixing $(\hat{f}^{\alpha}_{\beta\gamma},h_{\alpha})$ and getting the $4d$ equations of motion by solving the Bianchi identity, one can first identify the on-shell degrees of freedom, for example, $\Phi^{\tilde{\sigma}} \sim \Phi^{[a(s+n),b(s)]}$ in the twisted-adjoint representation of the higher spin algebra, then find the suitable $(\hat{f}^{\alpha}_{\beta\gamma},h_{\alpha})$ so that the Bianchi identity is satisfied for the arbitrary $\Phi^{\tilde{\sigma}}$. 
\begin{equation}\label{5rt7th}
	\{H_{c_{1} \cdots c_{n}},\\n=0,1,\cdots\}\;\cup \;\{R^{[a(s-1),b(s-1)]}_{ab;c_{1} \cdots c_{n}}, s=2,4,\cdots, n  =0,1,\cdots\}
\end{equation}
and $\{\Phi^{[a(s+n),b(s)]}, s=0,2,\cdots, n=0,1,\cdots\}$ have the same number of indices. With the $4d$ equations of motion imposed on (\ref{5rt7th}), the two may contain the same number of degrees of freedom. Written in terms of $\Phi^{\tilde{\sigma}}$, the unfolded equation becomes 
\begin{equation}\label{fxg3}
	d W^{\alpha} =\frac{1}{2}\bar{f}^{\alpha}_{\beta\gamma}(\Phi^{\tilde{\sigma}})W^{\beta}\wedge W^{\gamma},  \;\;\;\;\;\;\;\;\;\;  d \Phi^{\tilde{\alpha}}=F^{\tilde{\alpha}}_{\beta}(\Phi^{\tilde{\sigma}})  W^{\beta}.
\end{equation}
It remains to find $(\bar{f}^{\alpha}_{\beta\gamma},F^{\tilde{\alpha}}_{\beta})$ satisfying the Bianchi identity and also giving rise to the correct free theory limit\footnote{As is shown in appendix C, there are $(\bar{f}^{\alpha}_{\beta\gamma},F^{\tilde{\alpha}}_{\beta})$ satisfying the Bianchi identity but failing to give the correct free theory limit. It is unclear whether the two requirements can uniquely fix $(\bar{f}^{\alpha}_{\beta\gamma},F^{\tilde{\alpha}}_{\beta})$ (up to a field redefinition) or not.}. Vasiliev theory gives the elegant solution to this problem \cite{4df, 5gt, 67h}. By solving the $Z$ part of the Vasiliev equation order by order, one may finally get the required $(\bar{f}^{\alpha}_{\beta\gamma},F^{\tilde{\alpha}}_{\beta})$ \cite{7vhg}.

For supersymmetry, it is also possible to study the dynamics of the 0-form matter on group manifold with the fixed background such as the WZ model. The component expansion of the 0-form on superspace gives the spin $0$ and $1/2$ fields in $4d$. The allowed gauge transformation is the global super Poincare transformation, which is the diffeomorphism transformation on $\textbf{M}$ preserving the background. For higher spin theory, one can similarly consider the 0-form $H$ on $\textbf{M}$ with   
\begin{equation}
	d W_{0}^{\alpha} = \frac{1}{2}f^{\alpha}_{\beta\gamma}W_{0}^{\beta} \wedge W_{0}^{\gamma},\;\;\;\;\;\;\;\;\;\;	d H = H_{\alpha}W_{0}^{\alpha}. 
\end{equation}
$W_{0}^{\alpha}$ describes the background with the vanishing curvature. The system has the global HS symmetry. The component expansion of $H$ on $\textbf{M}$ gives the spin $s=0,2,\cdots$ fields $R^{s}_{a_{1}\cdots a_{s},b_{1}\cdots b_{s}}$. On the other hand, the linearized Vasiliev equation for the 0-forms on background $W_{0}^{\alpha}$ is
\begin{equation}
	d \Phi^{\tilde{\alpha}}= k^{\tilde{\alpha}}_{\beta\tilde{\gamma}}\Phi^{\tilde{\gamma}} W_{0}^{\beta},
\end{equation}
which is also invariant under the global HS transformation. $k^{\tilde{\alpha}}_{\beta\tilde{\gamma}}$ is the constant. With $\Phi\equiv\Phi^{[a(0),b(0)]}=H$, from $	d \Phi= k_{\beta\tilde{\gamma}}\Phi^{\tilde{\gamma}} W_{0}^{\beta}$, we have $H_{\beta}=k_{\beta\tilde{\gamma}}\Phi^{\tilde{\gamma}} $. $R^{s}_{a_{1}\cdots a_{s},b_{1}\cdots b_{s}}$ can then be taken as the Weyl tensor of the linearized HS theory. With the space extended from $AdS_{4}$ to $\textbf{M}$, 0-forms in the linearized Vasiliev theory get the interpretation as the derivatives of a single 0-form $H$ on $\textbf{M}$.

The rest of the paper is organized as follows. In Section $2$, we construct a symmetric space $M$ with the higher spin transformation group the isometry group. In Section $3$, we consider the theory with the local higher spin symmetry. The discussion and conclusion are given in Section $4$.

\section{Symmetric space from the higher spin algebra}

We will consider the minimal bosonic higher spin theory in $AdS_{4}$ with the coordinate $u^{\mu}$, $\mu=1,2,3,4$. The related HS algebra is $ho(1|2:[3,2])$ with the basis $\{t_{\alpha}\sim t_{A_{1}\cdots A_{s-1},B_{1}\cdots B_{s-1}}\}$ \cite{mi}. $ t_{A_{1}\cdots A_{s-1},B_{1}\cdots B_{s-1}}$ is in irreducible representations of $SO(3,2)$ characterized by two row rectangular Young tableaux, $A_{i},B_{i} =0, 1,2,3,4$, $s=2,4,\cdots$ 
\begin{eqnarray}\label{46g}
\nonumber && 	t_{A_{1}\cdots A_{s-1},B_{1}\cdots B_{s-1}}=t_{\{A_{1}\cdots A_{s-1}\},B_{1}\cdots B_{s-1}}=t_{A_{1}\cdots A_{s-1},\{B_{1}\cdots B_{s-1}\}},  \\&& t_{\{A_{1}\cdots A_{s-1},A_{s}\}B_{2}\cdots B_{s-1}}=0,\;\;\;\;\;\;\;t_{A_{1}\cdots A_{s-3}CC,}^{\;\;\;\;\;\;\;\;\;\;\;\;\;\;\;\;\;\;\;\;\; B_{1}\cdots B_{s-1}}=0. 
\end{eqnarray}
With $a_{i},b_{i} = 1,2,3,4$, basis of $ho(1|2:[3,2])$ can be rewritten as 
\begin{eqnarray}\label{labd}
\nonumber \{t_{\alpha}\}&=& \{t_{A_{1}\cdots A_{s-1},B_{1}\cdots B_{s-1}}\}
 \\  &=& \{t_{0\cdots 0,b_{1}\cdots b_{s-1}},t_{0 \cdots 0a_{1},b_{1}\cdots b_{s-1}},t_{0\cdots 0a_{1} a_{2},b_{1}\cdots b_{s-1}},\cdots, t_{a_{1}\cdots a_{s-1},b_{1}\cdots b_{s-1}}\}.
\end{eqnarray}
Let
\begin{equation}\label{ae}
	 \{t_{Q} \}= \{t_{0 \cdots 0a_{1},b_{1}\cdots b_{s-1}},t_{0\cdots 0a_{1} a_{2}a_{3},b_{1}\cdots b_{s-1}},\cdots, t_{a_{1}\cdots a_{s-1},b_{1}\cdots b_{s-1}}\} 
\end{equation}
be the basis of $a[E]$,
\begin{equation}\label{aee}
	\{t_{A} \}= \{t_{0 \cdots 0,b_{1}\cdots b_{s-1}},t_{0\cdots 0a_{1} a_{2},b_{1}\cdots b_{s-1}},\cdots, t_{0a_{1}\cdots a_{s-2},b_{1}\cdots b_{s-1}}\} 
\end{equation}
be the basis of $K$, $ho(1|2:[3,2])=a[E]\oplus K$. 
\begin{equation}\label{na}
	[a[E],a[E]]\subset a[E],\;\;\;\;\;\;\;\;\;[a[E],K]\subset K,\;\;\;\;\;\;\;\;\;[K,K]\subset a[E]. 
\end{equation}
$a[E]$ is a subalgebra of $ho(1|2:[3,2])$ generating a subgroup $E$. The coset space $G[ho(1|2:[3,2])]/E$ is a symmetric space according to (\ref{na}). With the group given, it is a standard procedure in mathematics to construct the group manifold $\textbf{M}$ for $G[ho(1|2:[3,2])]$\footnote{HS transformation group is the global symmetry group of the $3d$ $O(N)$ vector model and the dual minimal bosonic HS theory in $AdS_{4}$. The related algebra is $ho(1|2:[3,2])$. The group can be non-connected, just as $SO(3,1)$. Here $G[ho(1|2:[3,2])]$ refers to the the connected piece containing the identity, which is a simple group. So the related group manifold $\textbf{M}$ is also connected.} and the symmetric space $M$ for $G[ho(1|2:[3,2])]/E$. In the following, we will give a construction based on the operators and the conserved charges of the quantum higher spin theory in $AdS_{4}$. For earlier work on space with the tensor coordinates, see \cite{1z, gbfb9vn}.

In quantum higher spin theory, there are conserved charges $\{Q_{A_{1}\cdots A_{s-1},B_{1}\cdots B_{s-1}}\}$ in one-to-one correspondence with $\{t_{A_{1}\cdots A_{s-1},B_{1}\cdots B_{s-1}}\}$. In particular, $\{Q_{A_{1},B_{1}}\}$ are generators of $SO(3,2)$. Suppose $0$ is a point in the bulk of $AdS_{4}$, for example, $(1,0,0,0,0)$ in $x^{2}_{0}-x^{2}_{1}-x^{2}_{2}-x^{2}_{3}+ x^{2}_{4} = 1$, and $O(0)$ is the operator for the spin $0$ field at $0$, then the orbit generated by $SO(3,2)$ gives operators for the spin $0$ field in the entire $AdS_{4}$.
\begin{equation}
	\{O(u)|u \in AdS_{4}\}=\{g O(0)g^{-1}|g \in SO(3,2)\} ,
\end{equation}
where $g=e^{i\omega^{A_{1},B_{1}}Q_{A_{1},B_{1}}}$. Aside from $AdS_{4}$, the orbit generated by $G[ho(1|2:[3,2])]$ gives operators in an enlarged space $M$.
\begin{equation}\label{27}
	\{O(z)|z \in M\}=\{g O(0)g^{-1}|g \in G[ho(1|2:[3,2])]\} ,
\end{equation}
where $g=e^{i\omega^{A_{1}\cdots A_{s-1},B_{1}\cdots B_{s-1}}Q_{A_{1}\cdots A_{s-1},B_{1}\cdots B_{s-1}}}$. In $G[ho(1|2:[3,2])]$, there is a subgroup $E (z)$, $\forall\;e\in E(z)$, $e O(z) e^{-1} = O(z)$. The higher spin algebra is decomposed as 
\begin{equation}
	 ho(1|2:[3,2])=  K(z)\oplus a[E(z)] =g(z)K(0)g(z)^{-1}\oplus g(z) a[E(0)]g(z)^{-1}
\end{equation}
with $ K(z)$ the tangent space of $M$ at $z$. $M$ is the coset space $G[ho(1|2:[3,2])]/E $. In particular, $SO(3,1) \subset E$, $SO(3,2) \subset G[ho(1|2:[3,2])]$, $AdS_{4}=SO(3,2)/SO(3,1)$, so $M$ has a fiber bundle structure with the fiber $AdS_{4}$ attached at each point of the base manifold.

It remains to determine the subalgebra $a[E]$. Although the direct quantization of the higher spin theory in $AdS_{4}$ is still not available, its CFT dual is quite simple. In appendix A, the CFT realization of $O(0)$, or more accurately, $O^{+}(0)$, is given. It is shown that the charge $Q_{0\cdots 0a_{1} \cdots a_{2k-1},b_{1}\cdots b_{s-1}}$ corresponding to (\ref{ae}) commutes with $O(0)$. So $a[E]$ constructed here is the same as (\ref{ae}).

The metric on the coset space $M=G[ho(1|2:[3,2])]/E$ is defined in group theory. Alternatively, we can use the operator $O(z)$ to get the same result. There is a one-to-one correspondence between $T_{z}(M) = \{v^{M}\partial_{M}|M = 1,\cdots,\dim M \}$ and $K(z)$. For the given $\partial_{M}$, $\bar{\exists}\; k_{M}(z) \in K(z)$ satisfying
\begin{equation}
\partial_{M}O(z) = i[k_{M}(z),O(z)].  
\end{equation}
$\{k_{M}(z) \}$ compose the basis for $K(z)$, from which, one can define a special set of the coordinate on $M$
\begin{equation}\label{hu8}
	O(z) = e^{i k_{M}(0)z^{M}}O(0)e^{-i k_{M}(0)z^{M}}.
\end{equation}
The metric on $T_{z}(M)$ can be induced from $K(z)$, i.e.
\begin{equation}
	g_{MN}(z)  = \left\langle k_{M}(z) |k_{N}(z) \right\rangle,
\end{equation}
where $\left\langle k_{M}(z) |k_{N}(z) \right\rangle$ is the killing form. $g_{MN}$ is $G[ho(1|2:[3,2])]$ invariant. Under the $G[ho(1|2:[3,2])]$ transformation,
\begin{equation}
	O(z) \rightarrow g O(z)g^{-1}=O(z'). 
\end{equation}
$G[ho(1|2:[3,2])]$ generates the isometric transformation $z \rightarrow z'$ on $M$.

The tangent space on the coset space $M$ is $\{t_{A} \}$. The group manifold of $ho(1|2:[3,2])$ is the manifold $\textbf{M}$ with the tangent space $\{t_{\alpha}\}$, $\dim \textbf{M} = \dim ho(1|2:[3,2])$. The coordinate on $\textbf{M}$ is $Z_{\bar{M}}$, $i k_{\bar{M}}(Z)= \partial_{\bar{M}}g(Z)g(Z)^{-1}$, $G_{\bar{M}\bar{N}}(Z)  = \left\langle k_{\bar{M}}(Z) |k_{\bar{N}}(Z) \right\rangle$.
\begin{equation}
	\partial_{\bar{M}}O(Z) = i[k_{\bar{M}}(Z),O(Z)].
\end{equation}
When $k_{\bar{M}}(Z) \in E(Z)$, $\partial_{\bar{M}}O(Z)=0$. Let $\{t_{\alpha}\}$ be a set of the orthogonal normalized basis of $ho(1|2:[3,2])$, one may assume $k_{\alpha}(Z)=g(Z)t_{\alpha}g(Z)^{-1}$. $k_{\bar{M}}(Z) = W^{\alpha}_{\bar{M}}(Z)k_{\alpha}(Z)$ and $k_{\alpha}(Z)=W^{\bar{M}}_{\alpha}(Z)k_{\bar{M}}(Z)$ gives the vielbein on $\textbf{M}$:
\begin{equation}
W^{\alpha}_{\bar{M}}W^{\bar{M}}_{\beta}=\delta^{\alpha}_{\beta},\;\;\;\;\;W^{\alpha}_{\bar{M}}W^{\bar{N}}_{\alpha}=\delta^{\bar{N}}_{\bar{M}},\;\;\;\;\;\eta_{\alpha\beta}W^{\alpha}_{\bar{M}}W^{\beta}_{\bar{N}}=G_{\bar{M}\bar{N}}.
\end{equation}
$\eta_{\alpha\beta}=f^{\rho}_{\alpha\sigma}f^{\sigma}_{\beta\rho}=\left\langle t_{\alpha}|t_{\beta}\right\rangle$ is the killing metric for $ho(1|2:[3,2])$ with the suitable normalization assumed.\footnote{Notice that the killing metric $\eta_{\alpha\beta}$ is indefinite having one sign for compact directions and the opposite for non-compact directions. $G(ho(1|2:[3,2])$ is obviously not a compact group as one can see from its subgroup $SO(3,2)$.} Suppose $\partial_{\bar{N}}k_{\bar{M}}(Z) = \Gamma^{\bar{L}}_{\bar{N}\bar{M}}k_{\bar{L}}(Z)$, $\partial_{\bar{N}}k_{\alpha}(Z) = \phi^{\beta}_{\bar{N}\alpha}k_{\beta}(Z)$, there will be
\begin{equation}
\partial_{\bar{N}}W^{\bar{M}}_{\alpha}+\Gamma^{\bar{M}}_{\bar{N}\bar{L}}W^{\bar{L}}_{\alpha}-\phi^{\beta}_{\bar{N}\alpha}W^{\bar{M}}_{\beta}=0.	
\end{equation}
With the covariant derivative defined as $	\mathscr{D}_{\bar{M}}=\partial_{\bar{M}}-\Gamma_{\bar{M}}$ and $\mathscr{D}_{\alpha}=W^{\bar{M}}_{\alpha}(\partial_{\bar{M}}-\phi_{\bar{M}})=\partial_{\alpha}-\phi_{\alpha}$, we have
\begin{eqnarray}
 && 	\mathscr{D}_{\bar{M}_{n}}\cdots \mathscr{D}_{\bar{M}_{2}}\mathscr{D}_{\bar{M}_{1}}O(Z)=i^{n}[k_{\bar{M}_{1}}(Z),[k_{\bar{M}_{2}}(Z),\cdots[k_{\bar{M}_{n}}(Z),	O(Z)]\cdots]],\\
 && \mathscr{D}_{\alpha_{n}}\cdots \mathscr{D}_{\alpha_{2}}\mathscr{D}_{\alpha_{1}}O(Z)=i^{n}[k_{\alpha_{1}}(Z),[k_{\alpha_{2}}(Z),\cdots[k_{\alpha_{n}}(Z),	O(Z)]\cdots]].
\end{eqnarray}

As is shown in Appendix A, for $[Q_{0\cdots0 a_{1}\cdots a_{s},b_{1}\cdots b_{s+k}},O(0)]$ with $k=1,3,\cdots$  
\begin{eqnarray}
\nonumber && [Q_{0\cdots0 a_{1}\cdots a_{s},b_{1}\cdots b_{s+k}},O(0)]\\\nonumber
 &=& \sum^{t=1,2,\cdots,2s+k-2r}_{r=0,2,\cdots,s}g^{c_{1}\cdots c_{2r+t}}_{0\cdots0 a_{1}\cdots a_{s},b_{1}\cdots b_{s+k}} [Q_{0 c_{1}\cdots c_{r},c_{r+1}\cdots c_{2r+1}},\cdots [Q_{0,c_{2r+t-1}},[Q_{0,c_{2r+t}},O(0)]]\cdots]. \\
\end{eqnarray}
At the point $Z$, $O(Z) =g(Z) O(0)g(Z)^{-1}$, $Q_{A}(Z) =g(Z) Q_{A} g(Z)^{-1} $,
\begin{eqnarray}
\nonumber && [Q_{0\cdots0 a_{1}\cdots a_{s},b_{1}\cdots b_{s+k}}(Z),O(Z)]\\\nonumber
 &=& \sum^{t=1,2,\cdots,2s+k-2r}_{r=0,2,\cdots,s}g^{c_{1}\cdots c_{2r+t}}_{0\cdots0 a_{1}\cdots a_{s},b_{1}\cdots b_{s+k}} \\&& [Q_{0 c_{1}\cdots c_{r},c_{r+1}\cdots c_{2r+1}}(Z),\cdots [Q_{0,c_{2r+t-1}}(Z),[Q_{0,c_{2r+t}}(Z),O(Z)]]\cdots] .  
\end{eqnarray}
Since 
\begin{eqnarray}
\nonumber&& \mathscr{D}_{0,b_{s+k}}\mathscr{D}_{0,b_{s+k-1}}\cdots	\mathscr{D}_{0 a_{1}\cdots a_{s},b_{1}\cdots b_{s+1}}O(Z) \\&=& i^{k}  [Q_{0 a_{1}\cdots a_{s},b_{1}\cdots b_{s+1}}(Z),\cdots [Q_{0,b_{s+k-1}}(Z),[Q_{0,b_{s+k}}(Z),O(Z)]]\cdots],
\end{eqnarray}
there will be 
\begin{eqnarray}\label{D00d1}
\nonumber&& \partial_{0\cdots0 a_{1}\cdots a_{s},b_{1}\cdots b_{s+k}}O(Z) \\\nonumber&=& 
\sum^{t=1,2,\cdots,2s+k-2r}_{r=0,2,\cdots,s}i^{1-t}g^{c_{1}\cdots c_{2r+t}}_{0\cdots0 a_{1}\cdots a_{s},b_{1}\cdots b_{s+k}}  
\mathscr{D}_{0,c_{2r+t}}\mathscr{D}_{0,c_{2r+t-1}}\cdots	\mathscr{D}_{0 c_{1}\cdots c_{r},c_{r+1}\cdots c_{2r+1}}O(Z)
.\\
\end{eqnarray}
According to the definition, $\phi^{\beta}_{\bar{M} \gamma}$ and $W^{\bar{M}}_{\alpha}$ are invariant under the global higher spin transformation, so is their contraction $\phi_{\alpha}$. $\phi_{\alpha}$ is a scalar, so it must be a constant on $\textbf{M}$. (\ref{D00d1}) can be further rewritten as  
\begin{eqnarray}\label{D00d11}
\nonumber&& \partial_{0\cdots0 a_{1}\cdots a_{s},b_{1}\cdots b_{s+k}}O(Z) \\\nonumber&=& 
\sum^{t=1,2,\cdots,2s+k-2r}_{r=0,2,\cdots,s}G^{c_{1}\cdots c_{2r+t}}_{0\cdots0 a_{1}\cdots a_{s},b_{1}\cdots b_{s+k}}  
\partial_{0,c_{2r+t}}\partial_{0,c_{2r+t-1}}\cdots	\partial_{0 c_{1}\cdots c_{r},c_{r+1}\cdots c_{2r+1}}O(Z)
\\
\end{eqnarray}
for some constant $G^{c_{1}\cdots c_{2r+t}}_{0\cdots0 a_{1}\cdots a_{s},b_{1}\cdots b_{s+k}}$.

Just as the chiral constraint relates $\partial_{\bar{\theta}}$ with $\partial_{\mu}$, here, $\partial_{0\cdots0 a_{1}\cdots a_{s},b_{1}\cdots b_{s+k}}$ is determined by $\partial_{0,c_{2r+t}}\partial_{0,c_{2r+t-1}}\cdots	\partial_{0 c_{1}\cdots c_{r},c_{r+1}\cdots c_{2r+1}}$. This is because $[Q_{A_{1}\cdots A_{s-1},B_{1}\cdots B_{s-1}}(Z),O(Z)]\left|0\right\rangle$ are all in the 1-particle Hilbert space of the higher spin theory, for which
\begin{eqnarray}\label{ba1}
\nonumber&& \{[Q_{0,b_{s+k}}(Z),\cdots [Q_{0,b_{s+2}}(Z),[Q_{0 a_{1}\cdots a_{s},b_{1}\cdots b_{s+1}}(Z),O(Z)]]\cdots]\} \\&\sim & \{[Q_{0 a_{1}\cdots a_{s},b_{1}\cdots b_{s+1}}(Z),[Q_{0,b_{s+2}}(Z),\cdots ,[Q_{0,b_{s+k}}(Z),O(Z)]\cdots]]\}
\end{eqnarray}
compose the complete basis.\footnote{More precisely, it is $\{[Q_{0,b_{s+k}}(Z),\cdots [Q_{0,b_{s+2}}(Z),O^{s}_{ a_{1}\cdots a_{s},b_{1}\cdots b_{s}}(Z)]\cdots]\}$ that forms the complete basis, but (\ref{ba1}) is enough to generate $[Q_{A_{1}\cdots A_{s-1},B_{1}\cdots B_{s-1}}(Z),O(Z)]$ since $[Q_{Q}(Z),O(Z)]=0$ for $t_{Q}$ in (\ref{ae}).} In \cite{gbfb9vn}, by considering the zeroth level of the unfolded equation for the 0-form $\Phi$ in $M$, the similar result is also obtained. $\Phi=\Phi^{[a(0),b(0)]}$ is the lowest component of $\Phi^{[a(s+t),b(s)]}$. Generically, one may expect 
\begin{eqnarray}
\nonumber	&&[Q_{0\cdots0 a^{p}_{1}\cdots a^{p}_{s_{p}},b^{p}_{1}\cdots b^{p}_{s_{p}+k_{p}}}(Z),\cdots		[Q_{0\cdots0 a^{1}_{1}\cdots a^{1}_{s_{1}},b^{1}_{1}\cdots b^{1}_{s_{1}+k_{1}}}(Z),O(Z)]\cdots]   \\\nonumber& \sim &   \sum  \alpha (a_{1}\cdots a_{s},b_{1}\cdots b_{s+k})[Q_{0 a_{1}\cdots a_{s},b_{1}\cdots b_{s+1}}(Z),\cdots [Q_{0,b_{s+k-1}}(Z),[Q_{0,b_{s+k}}(Z),O(Z)]]\cdots],\\
\end{eqnarray}
where $\alpha (a_{1}\cdots a_{s},b_{1}\cdots b_{s+k})$ are constants to be determined. 
\begin{eqnarray}\label{D00d}
&&\nonumber	\partial_{0\cdots0 a^{1}_{1}\cdots a^{1}_{s_{1}},b^{1}_{1}\cdots b^{1}_{s_{1}+k_{1}}} \cdots		\partial_{0\cdots0 a^{p}_{1}\cdots a^{p}_{s_{p}},b^{p}_{1}\cdots b^{p}_{s_{p}+k_{p}}}O(Z) \\&\sim &   \sum  \Lambda (a_{1}\cdots a_{s},b_{1}\cdots b_{s+k})\partial_{0,b_{s+k}} \partial_{0,b_{s+k-1}}\cdots \partial_{0 a_{1}\cdots a_{s},b_{1}\cdots b_{s+1}}O(Z).
\end{eqnarray}
(\ref{D00d}) is the $G[ho(1|2:[3,2])]$-invariant differential operators on $M$, which will be useful in section 3.6 when we try to construct the theory with the global higher spin symmetry. .

\section{Theory with the local higher spin symmetry}

In section 2, the background in $\textbf{M}$ is fixed to be the intrinsic geometry with $d W_{0}^{\alpha} -\frac{1}{2}f^{\alpha}_{\beta\gamma}W_{0}^{\beta}\wedge W_{0}^{\gamma}=0$, which is invariant under the global higher spin transformation preserving $W_{0}^{\alpha}$. To have the local higher spin symmetry, the 1-form $W^{\alpha}$ in $\textbf{M}$ should be dynamical. We will study the dynamics of the 1-form $W^{\alpha}$ and the 0-form $H$ in $\textbf{M}$. With the suitable rheonomy condition and the torsion constraint imposed, $(W^{\alpha},H)$ in the whole $\textbf{M}$ is determined by $(W^{[a(s-1),b(0)]}_{\mu},H)$ in $AdS_{4}$. We then discuss the relation between the unfolded equation in group manifold approach and the unfolded equation in Vasiliev theory. We will also make a comment on theory with the global higher spin symmetry.

\subsection{Higher spin theory on group manifold and the rheonomy condition}

The 1-form $W^{\alpha}_{\bar{M}}$ is the vielbein on $\textbf{M}$. $W^{\alpha}_{\bar{M}}W^{\bar{M}}_{\beta}=\delta^{\alpha}_{\beta}$, $W^{\alpha}_{\bar{M}}W^{\bar{N}}_{\alpha}=\delta^{\bar{N}}_{\bar{M}}$, $\eta_{\alpha\beta}W^{\alpha}_{\bar{M}} W^{\beta}_{\bar{N}}=G_{\bar{M}\bar{N}}$.\footnote{Here, $W^{\alpha}_{\bar{M}}$ is invertible, which is general enough to account for the $4d$ HS theory, in which, the relevant field is $W^{\alpha}_{\mu}$. Let $\{\alpha\}=\{\tilde{\alpha}\}\cup\{a\}$, $\{\bar{M}\}=\{\tilde{M}\}\cup \{\mu\}$, $W^{a}_{\mu}\sim e^{a}_{\mu}$ is usually required to be invertible, one can also suitably select $W^{\tilde{\alpha}}_{\tilde{M}}$ to make the whole $W^{\alpha}_{\bar{M}}$ invertible.} The curvature 2-form is defined as 
\begin{equation}
R^{\alpha} = d W^{\alpha} -\frac{1}{2}f^{\alpha}_{\beta\gamma}W^{\beta}\wedge W^{\gamma}.   
\end{equation}
It is convenient to use the 0-form $R^{\alpha}_{\beta\gamma}$ to parameterize $R^{\alpha}_{\bar{M}\bar{N}}$, $R^{\alpha}_{\bar{M}\bar{N}} =R^{\alpha}_{\beta\gamma} W^{\beta}_{\bar{M}}W^{\gamma}_{\bar{N}} $, $R^{\alpha}_{\beta\gamma} = R^{\alpha}_{\bar{M}\bar{N}}  W^{\bar{M}}_{\beta}W^{\bar{N}}_{\gamma}$. 
\begin{equation}\label{4rtgxk}
	d W^{\alpha} =\frac{1}{2}(f^{\alpha}_{\beta\gamma}+ R^{\alpha}_{\beta\gamma})W^{\beta}\wedge W^{\gamma}=\frac{1}{2}\hat{f}^{\alpha}_{\beta\gamma}W^{\beta}\wedge W^{\gamma},
\end{equation}
where $\hat{f}^{\alpha}_{\beta\gamma}  $ is the deformed structure constant. The Bianchi identity is
\begin{equation}\label{4rtgxk1}
\partial_{[\gamma} \hat{f}^{\alpha}_{\rho\sigma]}+\hat{f}^{\alpha}_{\beta[\gamma}\hat{f}^{\beta}_{\rho\sigma]}=0,
\end{equation}
where $\partial_{\gamma}=W^{\bar{M}}_{\gamma}\partial_{\bar{M}}$. In addition, we can add the 0-form matter field $H$ on $\textbf{M}$, 
\begin{eqnarray}
 \label{45h} && d H = H_{\alpha} W^{\alpha} \;\; \Leftrightarrow \;\; \partial_{\alpha}H = H_{\alpha},
 \\ \label{4rtgxk11}&& \partial_{[\rho}H_{\sigma]}+H_{\alpha}\hat{f}^{\alpha}_{\rho\sigma} = 0.
\end{eqnarray} 
The group manifold $\textbf{M}$ is necessarily involved in the definition of $R^{\alpha}_{\beta\gamma}$ and $H_{\alpha}$. (\ref{4rtgxk1}) and (\ref{4rtgxk11}) are defined in $\textbf{M}$ as well.

The definition (\ref{4rtgxk}) and (\ref{45h}) is invariant under the diffeomorphism transformation generated by $\xi^{\bar{M}}$, 
\begin{equation}\label{4rt11}
	\delta_{\xi} W_{\bar{M}}^{\alpha}=\xi^{\bar{N}}  \partial_{\bar{N}}W_{\bar{M}}^{\alpha}+\partial_{\bar{M}}\xi^{\bar{N}}W_{\bar{N}}^{\alpha}, \;\;\;\;	\delta_{\xi}\hat{f}^{\alpha}_{\beta\gamma}=
\xi^{\bar{N}}  \partial_{\bar{N}}\hat{f}^{\alpha}_{\beta\gamma}, \;\;\;\;\delta_{\xi}H = \xi^{\bar{N}}\partial_{\bar{N}}H     , \;\;\;\; \delta_{\xi}H_{\alpha} = \xi^{\bar{N}}\partial_{\bar{N}}H_{\alpha}  .
\end{equation}
With
\begin{equation}
	\epsilon^{\alpha}=\xi^{\bar{M}}W^{\alpha}_{\bar{M}},\;\;\;\;\;\;\;\;\;\;\;\xi^{\bar{M}} = \epsilon^{\alpha} W^{\bar{M}}_{\alpha},
\end{equation}
(\ref{4rt11}) can be rewritten as
\begin{equation}\label{87huk}
		\delta_{\epsilon} W^{\alpha}= d\epsilon^{\alpha}+\hat{f}^{\alpha}_{\beta\gamma} \epsilon^{\beta}W^{\gamma},\;\;\;\;\delta_{\epsilon} \hat{f}^{\alpha}_{\rho\sigma}=\epsilon^{\beta}\partial_{\beta}\hat{f}^{\alpha}_{\rho\sigma},\;\;\;\;  \delta_{\epsilon} H=\epsilon^{\beta}H_{\beta}     ,\;\;\;\;  \delta_{\epsilon} H_{\alpha}=\epsilon^{\beta}\partial_{\beta}H_{\alpha}       ,
\end{equation}
which is the deformed local higher spin transformation.
\begin{equation}
	\delta_{\epsilon_{2}}\delta_{\epsilon_{1}}-	\delta_{\epsilon_{1}}\delta_{\epsilon_{2}}=\delta_{[\epsilon_{2},\epsilon_{1}]}, \;\;\;\;\;\;\;\;[\epsilon_{2},\epsilon_{1}]^{\alpha} = \hat{f}^{\alpha}_{\beta\gamma}\epsilon_{2}^{\gamma}\epsilon^{\beta}_{1}.
\end{equation}
The algebra is closed with the deformed structure constant $\hat{f}^{\alpha}_{\beta\gamma}$.

If for some $\Lambda$, $R^{\alpha}_{\Lambda \gamma}=0$, $\hat{f}^{\alpha}_{\Lambda \gamma}= f^{\alpha}_{\Lambda \gamma}$, the local gauge transformation generated by $\epsilon^{\Lambda}$ is undeformed. It is necessary to require $R^{\alpha}_{[a(1),b(1)] \gamma}\equiv R^{\alpha}_{(ab) \gamma}=0$ to make the local Lorentz transformation undeformed. Also, since $H$ is a scalar, $H_{(ab)}=0$ should hold so that $\delta_{\epsilon^{ab}} H= \epsilon^{ab}H_{(ab)}=0$. From (\ref{4rtgxk1}) and (\ref{4rtgxk11}), 
\begin{equation}
\delta_{\epsilon^{ab}} R^{\alpha}_{\rho\sigma} =\epsilon^{ab}\partial_{(ab)}R^{\alpha}_{\rho\sigma}=\epsilon^{ab}[ f^{\alpha}_{(ab)\beta}R^{\beta}_{\rho\sigma}	+f^{\beta}_{(ab)[\rho}R^{\alpha}_{\sigma]\beta}], \;\;\;\;\;\delta_{\epsilon^{ab}} H_{\alpha} =\epsilon^{ab}\partial_{(ab)}H_{\alpha}=-\epsilon^{ab}f^{\beta}_{(ab)\alpha}H_{\beta}.
\end{equation}
The evolution along the $(ab)$ direction is a local Lorentz transformation, so the group manifold $\textbf{M}$ effectively reduces to the coset space $\mathcal{M}=G[ho(1|2:[3,2])]/SO(3,1)$. Recall that in Section 2, we have discussed the coset space $M=G[ho(1|2:[3,2])]/E$. For $\textbf{M}$ to reduce to $M$, there must be $R^{\alpha}_{Q\gamma}=0$ so that the local gauge transformation generated by $\epsilon^{Q}$ is undeformed. However, at least in Vasiliev theory, $R^{\alpha}_{(ab) \gamma}=0$ is valid but $R^{\alpha}_{Q\gamma}=0$ does not necessarily hold.

When $\beta \neq (ab)$, $\partial_{\beta}\hat{f}^{\alpha}_{\rho\sigma}$ and $\partial_{\beta}H_{\alpha}$ cannot be uniquely determined by (\ref{4rtgxk1}) and (\ref{4rtgxk11}). Nevertheless, from (\ref{4rtgxk1}) and (\ref{4rtgxk11}), we have 
\begin{eqnarray}
  \label{5678u} && 		\partial_{\gamma} R^{\alpha}_{ab}=\partial_{[b} R^{\alpha}_{a]\gamma }+\hat{f}^{\alpha}_{\beta[\gamma}\hat{f}^{\beta}_{ba]}
 \\ \label{5678uuu}&& \partial_{\gamma} H_{a} =\partial_{a}H_{\gamma}+H_{\alpha}\hat{f}^{\alpha}_{a \gamma}
\end{eqnarray} 
with $H_{a} = \partial_{a}H$. $a$ represents the $[0,a]$ element of $ho(1|2:[3,2])$. Let 
\begin{equation}
	R^{\alpha}_{ab;c_{1}\cdots c_{n}} = \partial_{c_{n}}\cdots \partial_{c_{1}} R^{\alpha}_{ab},\;\;\;\;\;\;\;\;H_{c_{1}\cdots c_{n}} = \partial_{c_{n}}\cdots \partial_{c_{1}} H ,
\end{equation}
if
\begin{eqnarray}\label{4r67h}
  \nonumber && 	R^{\alpha}_{\beta\gamma}=r^{\alpha}_{\beta\gamma} (R^{\sigma}_{ab},R^{\sigma}_{ab;c_{1}},\cdots,H,H_{c_{1}},\cdots)
 \\ && H_{\gamma}=h_{\gamma} (R^{\sigma}_{ab},R^{\sigma}_{ab;c_{1}},\cdots,H,H_{c_{1}},\cdots)
\end{eqnarray} 
with $r^{\alpha}_{\beta\gamma}$ and $h_{\gamma}$ the polynomials of $R^{\sigma}_{ab}, R^{\sigma}_{ab;c_{1}}, \cdots, H, H_{c_{1}}, \cdots$ with the constant coefficients, then 
\begin{eqnarray}
\label{fxgfe} && 	\partial_{\gamma} R^{\alpha}_{ab}=\partial_{[b} R^{\alpha}_{a]\gamma }+\hat{f}^{\alpha}_{\beta[\gamma}\hat{f}^{\beta}_{ba]}=r^{\alpha}_{ab;\gamma} (R^{\sigma}_{ab},R^{\sigma}_{ab;c_{1}},\cdots,H,H_{c_{1}},\cdots) 
 \\ 
 && \partial_{\gamma} H_{a} =\partial_{a}H_{\gamma}+H_{\alpha}\hat{f}^{\alpha}_{a \gamma}=h_{a;\gamma} (R^{\sigma}_{ab},R^{\sigma}_{ab;c_{1}},\cdots,H,H_{c_{1}},\cdots)
\end{eqnarray} 
are also polynomials. Moreover, since 
\begin{equation}\label{144f}
	(\partial_{\beta}\partial_{\gamma}-\partial_{\gamma}\partial_{\beta})F = \hat{f}^{\alpha}_{\gamma\beta}\partial_{\alpha}F,
\end{equation}
\begin{eqnarray}
 && 	(\partial_{c}\partial_{\gamma}-\partial_{\gamma}\partial_{c})R^{\alpha}_{ab} = \hat{f}^{\sigma}_{\gamma c}\partial_{\sigma}R^{\alpha}_{ab}=\hat{f}^{\sigma}_{\gamma c}r^{\alpha}_{ab;\sigma}(R^{\beta}_{ab},R^{\beta}_{ab;c_{1}},\cdots,H,H_{c_{1}},\cdots),
 \\ && 	(\partial_{c}\partial_{\gamma}-\partial_{\gamma}\partial_{c})H_{a} = \hat{f}^{\sigma}_{\gamma c}\partial_{\sigma}H_{a}=\hat{f}^{\sigma}_{\gamma c}h_{a;\sigma}(R^{\beta}_{ab},R^{\beta}_{ab;c_{1}},\cdots,H,H_{c_{1}},\cdots),
\end{eqnarray} 
so
\begin{equation}
\partial_{\gamma}R^{\alpha}_{ab;c} =\partial_{\gamma}\partial_{c}R^{\alpha}_{ab} =\partial_{c}r^{\alpha}_{ab;\gamma}-\hat{f}^{\sigma}_{\gamma c}r^{\alpha}_{ab;\sigma}=r^{\alpha}_{ab;c \gamma} 
\end{equation}
and 
\begin{equation}
\partial_{\gamma}H_{ac}=\partial_{\gamma}H_{a;c} =\partial_{\gamma}\partial_{c}H_{a} =\partial_{c}h_{a;\gamma}-\hat{f}^{\sigma}_{\gamma c}h_{a;\sigma}=h_{a;c \gamma} =h_{ac \gamma} 
\end{equation}
are again polynomials. Subsequently, one can prove for $n=0,1,\cdots$, we have
\begin{eqnarray}\label{138h}
  \nonumber && 		\partial_{\gamma} R^{\alpha}_{ab;c_{1}\cdots c_{n}} = r^{\alpha}_{ab;c_{1}\cdots c_{n} \gamma}(R^{\beta}_{ab},R^{\beta}_{ab;c_{1}},\cdots,H,H_{c_{1}},\cdots),
 \\ && \partial_{\gamma} H_{c_{1}\cdots c_{n}} = h_{c_{1}\cdots c_{n} \gamma}(R^{\beta}_{ab},R^{\beta}_{ab;c_{1}},\cdots,H,H_{c_{1}},\cdots),
\end{eqnarray}
or equivalently, 
\begin{eqnarray}
  \nonumber && 	d R^{\alpha}_{ab;c_{1}\cdots c_{n}} = r^{\alpha}_{ab;c_{1}\cdots c_{n} \gamma}(R^{\beta}_{ab},R^{\beta}_{ab;c_{1}},\cdots,H,H_{c_{1}},\cdots)W^{\gamma},
 \\ && d H_{c_{1}\cdots c_{n}} = h_{c_{1}\cdots c_{n} \gamma}(R^{\beta}_{ab},R^{\beta}_{ab;c_{1}},\cdots,H,H_{c_{1}},\cdots)W^{\gamma},
\end{eqnarray}
where $r^{\alpha}_{ab;c_{1}\cdots c_{n} \gamma} $ and $h_{c_{1}\cdots c_{n} \gamma}$ are polynomials of $R^{\beta}_{ab}, R^{\beta}_{ab; c_{1}}, \cdots, H, H_{c_{1}}, \cdots$. Finally, we get the unfolded equation 
\begin{eqnarray}\label{3e56}
  \nonumber && 	d W^{\alpha} =\frac{1}{2}(f^{\alpha}_{\beta\gamma}+r^{\alpha}_{\beta\gamma})W^{\beta}\wedge W^{\gamma}, \\ \nonumber &&d R^{\alpha}_{ab;c_{1}\cdots c_{n}} = r^{\alpha}_{ab;c_{1}\cdots c_{n} \gamma}W^{\gamma},
 \\ && d H_{c_{1}\cdots c_{n}} = h_{c_{1}\cdots c_{n} \gamma}W^{\gamma}
,
\end{eqnarray} 
with $n=0,1,\cdots$. $r^{\alpha}_{\beta\gamma}$, $r^{\alpha}_{ab;c_{1}\cdots c_{n} \gamma}$ and $h_{c_{1}\cdots c_{n} \gamma}$ are functions of $R^{\beta}_{ab}, R^{\beta}_{ab; c_{1}}, \cdots, H, H_{c_{1}}, \cdots$. From $(R^{\beta}_{ab}, R^{\beta}_{ab; c_{1}}, \cdots, H, H_{c_{1}}, \cdots)$ at one point, $(W^{\alpha},H)$ on the whole $\textbf{M}$ can be determined up to a gauge transformation. (\ref{3e56}) is invariant under the local gauge transformation (\ref{87huk}) which can now be explicitly written as  
\begin{eqnarray}
  \nonumber && 	\delta_{\epsilon} W^{\alpha}= d\epsilon^{\alpha}+\hat{f}^{\alpha}_{\sigma\gamma} \epsilon^{\sigma}W^{\gamma}, \\ \nonumber &&
\delta_{\epsilon} R^{\alpha}_{ab;c_{1}\cdots c_{n}}=\epsilon^{\sigma}r^{\alpha}_{ab;c_{1}\cdots c_{n} \sigma},
\\&&  \delta_{\epsilon} H_{c_{1}\cdots c_{n}}=\epsilon^{\sigma}h_{c_{1}\cdots c_{n} \sigma}       .
\end{eqnarray}
$(R^{\beta}_{ab},R^{\beta}_{ab;c_{1}},\cdots,H,H_{c_{1}},\cdots)$ forms a complete higher spin multiplet.

(\ref{4r67h}) is the rheonomy condition in higher spin theory. This is the most generic rheonomy condition requiring that the curvature $(R^{\alpha}_{\beta\gamma},H_{\gamma})$ is determined by its inner components $(R^{\alpha}_{ab},H_{a})$ as well as their inner derivatives. The condition, together with the Bianchi identity, gives the unfolded equation. The rheonomy condition in supergravity (\ref{11}) is a special situation, in which, the dependence on the inner derivatives vanishes. Therefore, supergravity does not contain the higher derivative interactions.

The parameterization (\ref{4r67h}) should satisfy the Bianchi identity 
\begin{equation}\label{5ft789t}
\partial_{[\gamma} R^{\alpha}_{\rho\sigma]}+\hat{f}^{\alpha}_{\beta[\gamma}\hat{f}^{\beta}_{\rho\sigma]}=0, \;\;\;\;\;\; \partial_{[\gamma} H_{\beta]} +H_{\alpha}\hat{f}^{\alpha}_{ \gamma\beta}=0.
\end{equation}
With (\ref{138h}) and (\ref{4r67h}) plugged in (\ref{5ft789t}), we get 
\begin{equation}\label{yt}
	F^{\alpha}_{[\gamma \rho\sigma]}(R^{\beta}_{ab},R^{\beta}_{ab;c_{1}},\cdots,H,H_{c_{1}},\cdots)=0,\;\;\;\;\;\;\;\;\;\;F_{[\beta\gamma]}(R^{\beta}_{ab},R^{\beta}_{ab;c_{1}},\cdots,H,H_{c_{1}},\cdots)=0,
\end{equation}
where $F^{\alpha}_{[\gamma \rho\sigma]}$ and $F_{[\beta\gamma]}$ are also polynomials of $R^{\beta}_{ab},R^{\beta}_{ab;c_{1}},\cdots,H,H_{c_{1}},\cdots$. (\ref{yt}) gives the $4d$ equations of motion for $(R^{\beta}_{ab},R^{\beta}_{ab;c_{1}},\cdots,H,H_{c_{1}},\cdots)$. For the randomly selected function $(r^{\alpha}_{\beta\gamma},h_{\gamma})$, (\ref{yt}) only has the trivial solution $R^{\beta}_{ab}=H=0$. $(r^{\alpha}_{\beta\gamma},h_{\gamma})$ should be chosen to allow as many on-shell degrees of freedom as possible. In this sense, (\ref{yt}) determines both $(r^{\alpha}_{\beta\gamma},h_{\gamma})$ and the $4d$ equations of motion.

To guarantee the local Lorentz invariance, in (\ref{4r67h}), $R^{\alpha}_{(ab)\beta}=H_{(ab)}=0$. Since 
\begin{eqnarray}\label{mm}
  \nonumber && 	\partial_{(ab)}R^{\alpha}_{\rho\sigma} = \frac{\partial R^{\alpha}_{\rho\sigma}}{\partial R^{\beta}_{de;c_{1}\cdots c_{n}} }\partial_{(ab)}R^{\beta}_{de;c_{1}\cdots c_{n}} + \frac{\partial R^{\alpha}_{\rho\sigma}}{\partial H_{c_{1}\cdots c_{n}} }\partial_{(ab)}H_{c_{1}\cdots c_{n}}, \\ &&
	\partial_{(ab)}H_{\alpha} = \frac{\partial H_{\alpha}}{\partial R^{\beta}_{de;c_{1}\cdots c_{n}} }\partial_{(ab)}R^{\beta}_{de;c_{1}\cdots c_{n}} + \frac{\partial H_{\alpha}}{\partial H_{c_{1}\cdots c_{n}} }\partial_{(ab)}H_{c_{1}\cdots c_{n}},
\end{eqnarray}
where $\partial_{(ab)}R^{\alpha}_{\rho\sigma}$, $\partial_{(ab)}H_{\alpha} $, $\partial_{(ab)}R^{\beta}_{de;c_{1}\cdots c_{n}} $ and $\partial_{(ab)}H_{c_{1}\cdots c_{n}}$ are all standard local Lorentz transformations, the coefficients in $r^{\alpha}_{\rho\sigma}$ and $h_{\alpha}$ should be the Lorentz invariants. In fact, (\ref{mm}) are also included in (\ref{5ft789t}), so the Lorentz invariance of $r^{\alpha}_{\rho\sigma}$ and $h_{\alpha}$ is also the requirement of the Bianchi identity if $R^{\alpha}_{(ab)\beta}=H_{(ab)}=0$.

We only considered the equation (\ref{3e56}) on group manifold $\textbf{M}$, since in that space, the diffeomorphism transformation and the local gauge transformation are in one-to-one correspondence. As the universal property of the unfolded equation \cite{vas}, (\ref{3e56}) is well-defined in space $m$ with $\dim m \geq 4$. If $\dim m > \dim \textbf{M}$, different diffeomorphism transformations may be realized as the same gauge transformation, i.e. there are flat directions with $\xi^{\bar{M}} W^{\alpha}_{\bar{M}}=0$; if $\dim m < \dim \textbf{M}$, some gauge transformation does not have the diffeomorphism realization like that in $AdS_{4}$.

The initial value is $(R^{\alpha}_{ab},R^{\alpha}_{ab;c_{1}},\cdots,H,H_{c_{1}},\cdots)$ at one point, it is desirable to express it in terms of $(W_{\mu}^{\alpha},H)$ as well as its $4d$ derivatives at that point. 
\begin{eqnarray}\label{438}
  \nonumber &&  R^{\alpha}_{\mu\nu}=r^{\alpha}_{\beta\gamma}W^{\beta}_{\mu}W^{\gamma}_{\nu} \\ \nonumber&& 
 \partial_{\lambda}R^{\alpha}_{\mu\nu}=( \frac{\partial r^{\alpha}_{\beta\gamma}}{\partial R^{\sigma}_{ab,c_{1}\cdots c_{n}}}r^{\sigma}_{ab,c_{1}\cdots c_{n}\rho}+ \frac{\partial r^{\alpha}_{\beta\gamma}}{\partial H_{c_{1}\cdots c_{n}}}h_{c_{1}\cdots c_{n}\rho})W^{\rho}_{\lambda} W^{\beta}_{\mu}W^{\gamma}_{\nu} +r^{\alpha}_{\beta\gamma}\partial_{\lambda}W^{\beta}_{[\mu}W^{\gamma}_{\nu]}
\\ \nonumber&&  
\cdots \\ \nonumber&& H_{\mu}=h_{\alpha}W^{\alpha}_{\mu}
 \\ \nonumber&& \partial_{\lambda}	H_{\mu}=( \frac{\partial h_{\alpha}}{\partial R^{\sigma}_{ab,c_{1}\cdots c_{n}}}r^{\sigma}_{ab,c_{1}\cdots c_{n}\rho}+ \frac{\partial h_{\alpha}}{\partial H_{c_{1}\cdots c_{n}}}h_{c_{1}\cdots c_{n}\rho}) W^{\rho}_{\lambda}     W^{\alpha}_{\mu}+h_{\alpha}\partial_{\lambda}W^{\alpha}_{\mu}
 \\ && \cdots
\end{eqnarray} 
$r$ and $h$ are functions of $(R^{\alpha}_{ab;c_{1}\cdots c_{n}},H_{c_{1}\cdots c_{n}})$. In (\ref{438}), the unknowns are $(R^{\alpha}_{ab;c_{1}\cdots c_{n}},H_{c_{1}\cdots c_{n}})$, while the number of equations is the same as the number of the degrees of freedom of $(R^{\alpha}_{\mu\nu;\lambda_{1}\cdots \lambda_{n}},H_{\lambda_{1}\cdots \lambda_{n}})$, where $\mu,\nu=1,2,3,4$. (\ref{5678u}) and (\ref{5678uuu}) also impose constraints on the off-shell $(R^{\alpha}_{ab;c_{1}\cdots c_{n}}, H_{c_{1}\cdots c_{n}})$ to make it have the same number of degrees of freedom as $(R^{\alpha}_{\mu\nu;\lambda_{1}\cdots \lambda_{n}},H_{\lambda_{1}\cdots \lambda_{n}})$, so in principle, from (\ref{438}), $(R^{\alpha}_{ab;c_{1}\cdots c_{n}},H_{c_{1}\cdots c_{n}})$ can be solved in terms of $(W^{\alpha}_{\mu},\partial_{\nu_{1}}W^{\alpha}_{\mu}, \cdots, H,\partial_{\nu_{1}}H,\cdots )$.
\begin{eqnarray}\label{gvxfg1}
  \nonumber &&  	R^{\alpha}_{ab;c_{1}\cdots c_{n}} = g^{\alpha}_{ab;c_{1}\cdots c_{n}}(W^{\sigma}_{\mu},\partial_{\nu_{1}}W^{\sigma}_{\mu},\cdots, H,\partial_{\nu_{1}}H,\cdots ),
 \\ && H_{c_{1}\cdots c_{n}} = q_{c_{1}\cdots c_{n}}(W^{\sigma}_{\mu},\partial_{\nu_{1}}W^{\sigma}_{\mu},\cdots, H,\partial_{\nu_{1}}H,\cdots ).
\end{eqnarray}

The local gauge transformation of $(W^{\alpha}_{\mu},H)$ in $AdS_{4}$ is
\begin{eqnarray}\label{sxde3}
  \nonumber &&  		\delta_{\epsilon} W_{\mu}^{\alpha}= \partial_{\mu}\epsilon^{\alpha}+\hat{f}^{\alpha}_{\sigma\gamma}(R^{\beta}_{ab},R^{\beta}_{ab;c_{1}},\cdots,H,H_{c_{1}},\cdots) \epsilon^{\sigma}W_{\mu}^{\gamma},
 \\ && \delta_{\epsilon} H = \epsilon^{\sigma} h_{\sigma}(R^{\beta}_{ab},R^{\beta}_{ab;c_{1}},\cdots,H,H_{c_{1}},\cdots).
\end{eqnarray} 
With (\ref{gvxfg1}) plugged in (\ref{sxde3}), 
\begin{eqnarray}\label{ffff}
  \nonumber &&  		\delta_{\epsilon} W_{\mu}^{\alpha}= \partial_{\mu}\epsilon^{\alpha}+u^{\alpha}_{\sigma\gamma}(W^{\sigma}_{\mu},\partial_{\nu_{1}}W^{\sigma}_{\mu},\cdots, H,\partial_{\nu_{1}}H,\cdots ) \epsilon^{\sigma}W_{\mu}^{\gamma},
 \\ && \delta_{\epsilon} H = \epsilon^{\sigma} v_{\sigma}(W^{\sigma}_{\mu},\partial_{\nu_{1}}W^{\sigma}_{\mu},\cdots, H,\partial_{\nu_{1}}H,\cdots)
\end{eqnarray} 
gives the local gauge transformation rule of the matter-gravity coupled system $(W^{\alpha}_{\mu},H)$ in $AdS_{4}$.

Since
\begin{equation}\label{1285}
	(R^{\beta}_{ab},R^{\beta}_{ab;c_{1}},\cdots,H,H_{c_{1}},\cdots)\sim (W^{\sigma}_{\mu},\partial_{\nu_{1}}W^{\sigma}_{\mu},\cdots, H,\partial_{\nu_{1}}H,\cdots ), 
\end{equation}
$(W^{\alpha}, H)$ on the whole $\textbf{M}$ is determined by the on-shell $(W^{\sigma}_{\mu},\partial_{\nu_{1}}W^{\sigma}_{\mu},\cdots, H,\partial_{\nu_{1}}H,\cdots )$ at one point, or equivalently, the on-shell $(W^{\alpha}_{\mu}, H)$ in $AdS_{4}$. This is the rheonomy in higher spin theory. As is shown in Section 2, although the space is $\textbf{M}$ with the infinite dimension, the physical Hilbert space is still the same as the $4d$ higher spin theory. Imposing the rheonomy condition is a way to project out the physical degrees of freedom.

\subsection{Group manifold approach to supergravity}

In this subsection, we will give a review of the group manifold approach for supergravity \cite{1ggg,11gg,1g,2h}. Some modification is made so that supergravity is treated in the same way as the above discussed higher spin theory.

For $\mathcal{N}=1$ supergravity in $R^{3,1}$, the coordinate in group manifold is $(x^{\mu},x^{\mu\nu},\theta^{\chi})$, the associated 1-form is $\nu^{A}= (\omega^{ab},e^{a},\psi^{\alpha})$~\footnote{Here, $\alpha$ is the spinor index and should be distinguished from $\alpha$ in the rest sections, which represents the adjoint representation of $ho(1|2:[3,2])$. Also, $\alpha$ here is equivalent to the spinor index $(\alpha,\dot{\alpha})$ in section 3.}, and the 0-form matter field is $H$. We have 
\begin{eqnarray}
   \label{h6g} &&d	\nu^{A}=\frac{1}{2} \hat{f}^{A}_{BC}\nu^{B} \wedge \nu^{C},\;\;\;\;\;\; dH = H_{A}\nu^{A},
\\ \label{90j} && \partial_{[E} \hat{f}^{A}_{BC]}+\hat{f}^{A}_{D[E}\hat{f}^{D}_{BC]}=0,\;\;\;\;\;\; 
\partial_{[A}H_{B]}+H_{C}\hat{f}^{C}_{AB}=0,
\end{eqnarray}
where $\hat{f}^{A}_{BC} =f^{A}_{BC} +R^{A}_{BC}$, $H_{B} = (H_{a},H_{(ab)},H_{\alpha})$. $f^{A}_{BC}$ is the structure constant of the super Poincare group $\overline{Osp(4|1)}$. (\ref{h6g}) is invariant under the diffeomorphism transformation in group manifold generated by $\xi^{\bar{M}}=(\xi^{\mu\nu},\xi^{\mu},\xi^{\chi})$
\begin{equation}\label{4rt116t}
	\delta_{\xi} \nu_{\bar{M}}^{A}=\xi^{\bar{N}}  \partial_{\bar{N}}\nu_{\bar{M}}^{A}+\partial_{\bar{M}}\xi^{\bar{N}}\nu_{\bar{N}}^{A}, \;\;\;\;	\delta_{\xi}\hat{f}^{A}_{BC}=
\xi^{\bar{N}}  \partial_{\bar{N}}\hat{f}^{A}_{BC}, \;\;\;\;  \delta_{\xi}H=
\xi^{\bar{N}}  \partial_{\bar{N}}H       , \;\;\;\;  \delta_{\xi}H_{A}=
\xi^{\bar{N}}  \partial_{\bar{N}}H_{A}     ,
\end{equation}
which, when written in terms of the components, are local Lorentz transformation, the $4d$ diffeomorphism transformation and the supersymmetry transformation respectively. With $\epsilon^{A}=\xi^{\bar{M}}\nu^{A}_{\bar{M}}$, (\ref{4rt116t}) can be rewritten as
\begin{equation}
		\delta_{\epsilon} \nu^{A}= d\epsilon^{A}+\hat{f}^{A}_{BC} \epsilon^{B}\nu^{C},\;\;\;\; \delta_{\epsilon} \hat{f}^{A}_{BC}=\epsilon^{D}  \partial_{D}\hat{f}^{A}_{BC},\;\;\;\;  \delta_{\epsilon} H=\epsilon^{D}  H_{D}  ,\;\;\;\;   \delta_{\epsilon} H_{A}=\epsilon^{D}  \partial_{D}H_{A} .
\end{equation}

Until now, no dynamics is involved at all. The dynamical information is brought by imposing the suitable constraints on $R^{A}_{BC}$ and $H_{A}$. Here, the constraints that will be imposed are

(a) Factorization condition $R^{A}_{(ab)C}=0=H_{(ab)}$;

(b) Rheonomy condition and the torsion constraint: 
\begin{eqnarray}
    (i)\;\; R^{A}_{BC} &=& r^{A}_{BC}(R^{cd}_{ab},R^{cd}_{ab,c_{1}},\cdots,R^{\alpha}_{ab},R^{\alpha}_{ab,c_{1}},\cdots,H,H_{c_{1}},\cdots,H_{\alpha},H_{\alpha;c_{1}}\cdots), \label{hy}\\ \nonumber  \textnormal{or} \;\;\;\;\;\;\;\;\;\;\;\;\;&& \;
\\ \nonumber  (ii)\;\;R^{A}_{BC}&=&r^{A}_{BC}(R^{cd}_{ab},R^{cd}_{ab,c_{1}},\cdots,R^{\alpha}_{ab},R^{\alpha}_{ab,c_{1}},\cdots,H,H_{c_{1}},\cdots)
 \\ H_{A} &=&h_{A}  (R^{cd}_{ab},R^{cd}_{ab,c_{1}},\cdots,R^{\alpha}_{ab},R^{\alpha}_{ab,c_{1}},\cdots,H,H_{c_{1}},\cdots).
\end{eqnarray} 
(a) is imposed so that the local Lorentz transformation is undeformed. In (b), the rheonomy condition requires that the lower index of the independent fields can only contain $a$ so that the whole dynamics in group manifold is determined by that in a $4d$ submanifold; torsion constraint requires that the upper index cannot be $a$ so that $\omega^{ab}$ can be solved in terms of the rest fields. There are two possibilities. In $(i)$, the final dynamical fields are $(e^{a}_{\mu},\psi^{\alpha}_{\mu},H,H_{\alpha})$ in $M_{4}$, which is the situation for $\mathcal{N}=1$ supergravity coupled to the WZ matter. In $(ii)$, the dynamical fields are $(e^{a}_{\mu},\psi^{\alpha}_{\mu},H)$ in $M_{4}$ like that in higher spin theory.

$r^{A}_{BC}$ and $h_{A}$ are polynomials, the coefficients of which should be selected so that some scaling relation is respected \cite{1g}. The weight of $t^{A}$ is denoted as $w(A)$, $w(a)=1$, $w(ab)=0$, $w(\alpha)=1/2$. The super Poincare algebra $[t_{A_{1}},t_{A_{2}}]=if^{A_{3}}_{A_{1}A_{2}}t_{A_{3}}$ is invariant under 
\begin{equation}
	t_{A_{i}}\rightarrow v^{-w(A_{i})}t_{A_{i}}
\end{equation}
The 0-forms $H_{A}$ and $R^{A}_{BC}$ have the weight $-w(A)$ and $w(A)-w(B)-w(C)$ as follows

$\;$

\begin{tabular}{ l c c c c c c c c  c r }
  $H_{a}$ & $H_{\alpha}$ & $R^{cd}_{ab}$& $R^{cd}_{a\alpha}$& $R^{cd}_{\alpha\beta}$& $R^{c}_{ab}$& $R^{c}_{a\alpha}$& $R^{c}_{\alpha\beta}$& $R^{\gamma}_{ab}$& $R^{\gamma}_{a\alpha}$& $R^{\gamma}_{\alpha\beta}$ \\
  $-1$ & $-\frac{1}{2}$& $-2$& $-\frac{3}{2}$& $-1$&$ -1$&$ -\frac{1}{2}$& $0$& $-\frac{3}{2}$& $-1$& $-\frac{1}{2} $\\
\end{tabular}

$\;$
\\
Especially, $(R^{cd}_{ab},R^{\alpha}_{ab},H,H_{a},H_{\alpha})$ have the weight $(-2,-3/2,0,-1,-1/2)$. $(ii)$ cannot satisfy the scaling relation thus should be ruled out. For $(i)$, with the $H_{A}$ odd terms dropped, the most general form of $r^{A}_{BC}$ is 
\begin{eqnarray}\label{180ii}
  \nonumber &&  R^{bc}_{a\alpha}=r^{bc}_{a\alpha}|^{de}_{\beta}R^{\beta}_{de}+r^{bc}_{a\alpha}|^{\beta ,d}H_{\beta }H_{d}, \\ \nonumber&& R^{c}_{ab}= r^{c}_{ab}|^{\alpha,\beta}H_{\alpha}H_{\beta},
\\ \nonumber&&  R^{\beta}_{a \alpha}=r^{\beta}_{a \alpha}|^{\rho,\sigma}H_{\rho}H_{\sigma},
 \\ \nonumber&& R^{cd}_{ \alpha \beta}= r^{cd}_{ \alpha \beta}|^{\rho,\sigma}H_{\rho}H_{\sigma},
 \\&& R^{c}_{a \alpha}=R^{\gamma}_{ \alpha\beta}=R^{c}_{\alpha\beta}=0, 
\end{eqnarray} 
where $r^{\ast}_{\ast\ast}|^{\ast\ast}_{\ast}=r^{\ast}_{\ast\ast}|^{\ast\ast}_{\ast}(H)$ are functions of $H$ since $H$ has the weight $0$. $r^{\ast}_{\ast\ast}|^{\ast\ast}_{\ast}$ should be a Lorentz invariant to preserve the local Lorentz invariance. Although the torsion constraint is also imposed, $R^{a}_{AB}$ does not need to vanish, see for example \cite{R}. However, if $H_{\alpha}=H_{a}=0$, $R^{a}_{AB}=0$, so in pure supergravity case, we do have $R^{a}_{AB}=0$. Due to the scaling relation, the rheonomy condition is greatly simplified. For supergravity in $AdS_{4}$ with the symmetry group $Osp(4|1)$, a constant $L$ with the weight $1$ is involved. $L\rightarrow \infty $ gives the flat space limit, so only the $L^{-n}$ terms with $n\geq  0$ are allowed in rheonomy condition. (\ref{180ii}) remains valid.

(\ref{180ii}) should satisfy the Bianchi identity 
\begin{equation}\label{mei}
	\partial_{[E} R^{A}_{BC]}+f^{A}_{D[E}R^{D}_{BC]}+R^{A}_{D[E}f^{D}_{BC]}+R^{A}_{D[E}R^{D}_{BC]}=0,\;\;\;\;\;\; 
\partial_{[A}H_{B]}+H_{C}f^{C}_{AB}+H_{C}R^{C}_{AB}=0.
\end{equation}
In pure supergravity situation with $H=0$, $r^{A}_{BC}$ becomes
\begin{eqnarray}
  \nonumber && R^{bc}_{a\alpha}=r^{bc}_{a\alpha}|^{de}_{\beta}R^{\beta}_{de},\\&& R^{c}_{ab}=  R^{\beta}_{a \alpha}= R^{cd}_{ \alpha \beta}=R^{c}_{a \alpha}= R^{\gamma}_{ \alpha\beta}=R^{c}_{ \alpha\beta}=0.
\end{eqnarray}
(\ref{mei}) reduces to 
\begin{eqnarray}
 &&		\partial_{(ab)} R^{A}_{BC}=f^{A}_{(ab)D}R^{D}_{BC}-f^{D}_{(ab)C}R^{A}_{BD}-f^{D}_{(ab)B}R^{A}_{DC},\label{1}\\
 &&		f^{a}_{(ef)[b}R^{ef}_{cd]}=0 ,  \;\;\; \partial_{[a} R^{\alpha}_{bc]}=0 , \;\;\;  \partial_{[c} R^{ab}_{de]}+R^{ab}_{\alpha[c}R^{\alpha}_{de]}=0 ,\label{2}\\&&		\partial_{\beta} R^{\alpha}_{bc}+f^{\alpha}_{(ad)\beta}R^{(ad)}_{bc}=0
  ,\;\;\;  \partial_{\alpha} R^{ab}_{cd}+\partial_{[c} R^{ab}_{d]\alpha}=0 ,\label{3} \\ &&	f^{a}_{\beta\alpha}R^{\beta}_{bc}+f^{a}_{(ef)[b}R^{ef}_{c]\alpha}=0 ,  \;\;\;
 R^{\alpha}_{ac}f^{a}_{\beta\gamma }+ f^{\alpha}_{(ab)[\beta}R^{ab}_{ \gamma]c}=0
  ,    \;\;\; 
 \partial_{[\alpha} R^{ab}_{\beta]c}+R^{ab}_{dc}f^{d}_{\alpha\beta}=0  . \label{4}  
\end{eqnarray}
(\ref{1}) gives the Lorentz transformation of $R^{A}_{BC}$, which can be preserved in $r^{A}_{BC}(R^{ad}_{bc},R^{\beta}_{bc})$ if $r^{bc}_{a\alpha}|^{de}_{\beta}$ is a Lorentz scalar. (\ref{2}) are Bianchi identities in $4d$. (\ref{3}) gives the evolution of $(R^{ad}_{bc},R^{\beta}_{bc})$ along the $\alpha$ direction. With (\ref{3}) plugged in (\ref{4}), $r^{bc}_{a \alpha}|^{de}_{\beta}$ can be fixed and the $4d$ equations of motion 
\begin{equation}\label{vcfr}
	R^{cb}_{ab}-\frac{1}{2}\delta^{c}_{a}	R^{db}_{db}=0,\;\;\;\;\;\;\; \varepsilon^{abcd}(\gamma_{5}\gamma_{b})^{\alpha}_{\beta}R^{\beta}_{cd}=0
\end{equation}
come out. If we use the on-shell $\tilde{R}^{cb}_{ab}$ and $\tilde{R}^{\beta}_{cd}$ satisfying (\ref{vcfr}) to parameterize $r^{A}_{BC}$, (\ref{4}) will hold automatically. This is in analogy with Vasiliev theory, with $R^{\alpha}_{\beta\gamma}$ parametrized by the 0-form $\Phi^{\tilde{\alpha}}$ in the twisted-adjoint representation of the higher spin algebra, the Bianchi identity is satisfied for the arbitrary $\Phi^{\tilde{\alpha}}$.

Written as the unfolded equation,
\begin{eqnarray}\label{nanal}
  \nonumber && d \nu^{A}=\frac{1}{2}(f^{A}_{BC}+r^{A}_{BC})\nu^{B}\wedge\nu^{C} ~,~\\ \nonumber&& d R^{cd}_{ab;c_{1}\cdots c_{n}} =r^{cd}_{ab;c_{1}\cdots c_{n}A}\nu^{A} ~,~\;\;\;\;\;\;\;\;\;\;	d R^{\alpha}_{ab;c_{1}\cdots c_{n}} =r^{\alpha}_{ab;c_{1}\cdots c_{n}A}\nu^{A}~,~\\&& d H_{c_{1}\cdots c_{n}} =h_{c_{1}\cdots c_{n}A}\nu^{A} ~,~\;\;\;\;\;\;\;\;\;\;d H_{\alpha;c_{1}\cdots c_{n}} =h_{\alpha;c_{1}\cdots c_{n}A}\nu^{A} ~,~
\end{eqnarray}
where $r$, $h$ are all determined by $r^{A}_{BC}=r^{A}_{BC}(R^{cd}_{ab},R^{\alpha}_{ab},H_{a},H_{\alpha})$ and are functions of $(R^{cd}_{ab;c_{1}\cdots c_{n}},\\R^{\alpha}_{ab;c_{1}\cdots c_{n}},H_{c_{1}\cdots c_{n}},H_{\alpha;c_{1}\cdots c_{n}})$. With the on-shell $(R^{cd}_{ab;c_{1}\cdots c_{n}},R^{\alpha}_{ab;c_{1}\cdots c_{n}},H_{c_{1}\cdots c_{n}},H_{\alpha;c_{1}\cdots c_{n}})$ given at one point, $(\nu^{A},R^{cd}_{ab;c_{1}\cdots c_{n}},R^{\alpha}_{ab;c_{1}\cdots c_{n}},H_{c_{1}\cdots c_{n}},H_{\alpha;c_{1}\cdots c_{n}})$ on the whole $\textbf{M}$ can be solved. The local gauge transformation is
\begin{eqnarray}
  \nonumber && \delta_{\epsilon}\nu^{A}=d\epsilon^{A}+\hat{f}^{A}_{BC}\epsilon^{B}\nu^{C}, \\ \nonumber && \delta_{\epsilon} R^{cd}_{ab;c_{1}\cdots c_{n}} = \epsilon^{A}  r^{cd}_{ab;c_{1}\cdots c_{n}A}, \;\;\;\;\;\;\;\;\; \delta_{\epsilon} R^{\alpha}_{ab;c_{1}\cdots c_{n}} = \epsilon^{A}  r^{\alpha}_{ab;c_{1}\cdots c_{n}A},
\\  && \delta_{\epsilon} H_{c_{1}\cdots c_{n}} = \epsilon^{A}  h_{c_{1}\cdots c_{n}A},
\;\;\;\;\;\;\;\;\;\;\;\;\;\;\;  \delta_{\epsilon} H_{\alpha;c_{1}\cdots c_{n}} = \epsilon^{A}  h_{\alpha;c_{1}\cdots c_{n}A}.
\end{eqnarray} $(R^{cd}_{ab},R^{cd}_{ab;c_{1}},\cdots,R^{\alpha}_{ab},R^{\alpha}_{ab;c_{1}},\cdots,H,H_{c_{1}},\cdots,H_{\alpha},H_{\alpha;c_{1}},\cdots)$ compose the complete supersymmetry multiplet.

In addition to the 1-form $\nu^{A}$, the 0-form multiplet is introduced, forming the representation of the deformed local super Poincare transformation. The physical interpretation of the 0-form is the curvature and the matter field plus their derivatives. This is in the same spirit as the higher spin theory. Different from the higher spin theory, rheonomy condition (\ref{180ii}) only contains $R^{cd}_{ab},R^{\alpha}_{ab},H_{a},H_{\alpha}$, so the infinite length $0$-form multiplet does not enter into the $4d$ equations of motion. As a result, the equations of motion for $(e_{\mu}^{a},\psi_{\mu}^{\alpha},H,H_{\alpha})$ do not contain the higher order derivatives. One may similarly make a robust requirement $R^{\alpha}_{\beta\gamma}=r^{\alpha}_{\beta\gamma}(R^{\sigma}_{ab},H)$ and $H_{\gamma}=h_{\gamma}(R^{\sigma}_{ab},H)$ in higher spin theory. However, such $(r^{\alpha}_{\beta\gamma},h_{\gamma})$ may only allow the trivial solution $R^{\sigma}_{ab}=H=0$ when the Bianchi identity is imposed, no matter how coefficients in $(r^{\alpha}_{\beta\gamma},h_{\gamma})$ are adjusted.

Again,
\begin{eqnarray}\label{201}
  \nonumber && (R^{cd}_{ab},R^{cd}_{ab;c_{1}},\cdots,R^{\alpha}_{ab},R^{\alpha}_{ab;c_{1}},\cdots,H,H_{c_{1}},\cdots,H_{\alpha},H_{\alpha;c_{1}},\cdots) \\  &\sim & (e^{b}_{\mu},\partial_{\nu_{1}}e^{b}_{\mu},\cdots,\psi^{\alpha}_{\mu},\partial_{\nu_{1}}\psi^{\alpha}_{\mu},\cdots,H,\partial_{\nu_{1}}H,\cdots,H_{\alpha},\partial_{\nu_{1}}H_{\alpha},\cdots). 
\end{eqnarray}
With the on-shell $(e_{\mu}^{a},\psi_{\mu}^{\alpha},H,H_{\alpha})$ given on $M_{4}$, $(\nu^{A},H)$ on the whole group manifold can be determined up to a gauge transformation.

The dynamics is entirely encoded in function $r^{A}_{BC}(R^{ad}_{bc},R^{\beta}_{bc},H,H_{\alpha})$. By setting $H$ to $0$, we obtain the pure supergravity situation. Alternatively, one can consider the dynamics of the 0-form matter on the fixed supergravity background by setting $r^{A}_{BC}$ to $0$. With $\hat{f}^{A}_{BC}=f^{A}_{BC}$, (\ref{h6g}) and (\ref{90j}) reduce to
\begin{equation}\label{5tfg}
d	\nu_{0}^{A}=\frac{1}{2} f^{A}_{BC}\nu_{0}^{B} \wedge \nu_{0}^{C},\;\;\;\;\;\; dH = H_{A}\nu_{0}^{A},\;\;\;\;\;\; \partial_{[A}H_{B]}+H_{C}f^{C}_{AB}=0.
\end{equation}
$\nu_{0}^{A}$ describes the intrinsic geometry of the group manifold. The allowed gauge transformation parameter $\epsilon_{0}^{A}$ should make $\nu_{0}^{A}$ invariant
\begin{equation}\label{161sd}
	\delta_{\epsilon_{0}} \nu_{0\; \bar{M}}^{A}=\partial_{\bar{M}} \epsilon_{0}^{A} +f^{A}_{BC}\epsilon_{0}^{B}\nu^{C}_{0\; \bar{M}}=0.  
\end{equation}
$\epsilon_{0}^{A}$ generates the global super Poincare transformation on group manifold. 
\begin{equation}
	\delta_{\epsilon_{0}} H=\xi_{0}^{\bar{M}}\partial_{\bar{M}}H=\epsilon_{0}^{D}H_{D},\;\;\;\;\;\;\;\;\delta_{\epsilon_{0}} H_{A}=\xi_{0}^{\bar{M}}\partial_{\bar{M}}H_{A}=\epsilon_{0}^{D}\partial_{D}H_{A}.
\end{equation}
$\xi_{0}^{\bar{M}}\nu^{D}_{0 \; \bar{M}} = \epsilon^{D}_{0}$. Still, $H_{(ad)}=0$.   
\begin{eqnarray}
  \nonumber && \partial_{(ad)}H=0,\\\nonumber && 	\partial_{(ad)}H_{c}+H_{b}f^{b}_{(ad)c}=0,\\ && \partial_{(ad)}H_{\alpha}+H_{\beta}f^{\beta}_{(ad)\alpha}=0.
\end{eqnarray}
Evolution along $(ad)$ direction is a Lorentz transformation. One cannot assume $H_{\alpha}$ is the function of $(H,H_{c_{1}},H_{c_{1}c_{2}},\cdots)$, since the scaling relation is not respected. Let $\alpha=(\lambda,\dot{\lambda})$, one can at most require $H_{\dot{\lambda}}=0$, which is the chiral constraint for superfield.

\subsection{Imposing the torsion constraint in higher spin theory}

Back to higher spin theory, a further reduction of (\ref{1285}) can be made by imposing the following torsion constraint
\begin{eqnarray}\label{vgyh}
  \nonumber &&  	R^{\alpha}_{\beta\gamma}=r^{\alpha}_{\beta\gamma} (R^{[a(s-1),b(s-1)]}_{ab},R^{[a(s-1),b(s-1)]}_{ab;c_{1}},\cdots,H,H_{c_{1}},\cdots),  \\   & & H_{\gamma}=h_{\gamma}(R^{[a(s-1),b(s-1)]}_{ab},R^{[a(s-1),b(s-1)]}_{ab;c_{1}},\cdots,H,H_{c_{1}},\cdots). 
\end{eqnarray}
Namely, in (\ref{4r67h}), $\sigma$ is restricted to $[a(s-1),b(s-1)]$ with $s=2,4,\cdots$. In (\ref{438}), the number of equations is equal to the number of degrees of freedom of $(R^{\alpha}_{ab,c_{1}\cdots c_{n}},H_{c_{1}\cdots c_{n}})$ but the number of unknowns is equal to the degrees of freedom of $(R^{[a(s-1),b(s-1)]}_{ab,c_{1}\cdots c_{n}},H_{c_{1}\cdots c_{n}})$ now, so effectively, there will be some constraints imposed on $(W^{\alpha}_{\mu},H)$ in $AdS_{4}$ whose number is equal to the degrees of freedom of $R^{[a(s-1),b(t)]}_{ab}$ with $0\leq t \leq s-2$. It is expected that by solving these constraints, $W^{[a(s-1),b(t+1)]}_{\mu}$ can be expressed in terms of $(W^{[a(s-1),b(0)]}_{\mu},H)$. In fact, at least in free theory limit, imposing the torsion constraint $R^{[a(s-1),b(t)]}_{ab}=0$ for $0\leq t \leq s-2$ can indeed make $W^{[a(s-1),b(t+1)]}_{\mu}$ solved in terms of $W^{[a(s-1),b(0)]}_{\mu}$ \cite{free}. (\ref{1285}) then reduces to
\begin{eqnarray}\label{3e567}
  \nonumber && 	(R^{[a(s-1),b(s-1)]}_{ab},R^{[a(s-1),b(s-1)]}_{ab;c_{1}}, \cdots,H,H_{c_{1}},\cdots)   \\ &\sim &(W^{[a(s-1),b(0)]}_{\mu},\partial_{\nu_{1}}W^{[a(s-1),b(0)]}_{\mu},\cdots,H,\partial_{\nu_{1}}H,\cdots).
\end{eqnarray}
With $W^{[a(s-1),b(t+1)]}_{\mu}$ written in terms of $(W^{[a(s-1),b(0)]}_{\mu},H)$, (\ref{ffff}) becomes
\begin{eqnarray}\label{GAU}
  \nonumber \delta_{\epsilon} W_{\mu}^{[a(s-1),b(0)]}&=& 	 \partial_{\mu}\epsilon^{[a(s-1),b(0)]}+\epsilon^{\sigma}m^{[a(s-1),b(0)]}_{\sigma\gamma}(W^{[a(r-1),b(0)]}_{\mu},\partial_{\nu_{1}}W^{[a(r-1),b(0)]}_{\mu},\cdots,\\  \nonumber & &	  H,\partial_{\nu_{1}}H,\cdots )   w_{\mu}^{\gamma}(W^{[a(r-1),b(0)]}_{\mu},\partial_{\nu_{1}}W^{[a(r-1),b(0)]}_{\mu},\cdots,H,\partial_{\nu_{1}}H,\cdots ),\\ \delta_{\epsilon} H &=& \epsilon^{\sigma} n_{\sigma}(W^{[a(r-1),b(0)]}_{\mu},\partial_{\nu_{1}}W^{[a(r-1),b(0)]}_{\mu},\cdots, H,\partial_{\nu_{1}}H,\cdots ), 
\end{eqnarray}
which is the local gauge transformation rule of $(W^{[a(s-1),b(0)]}_{\mu},H)$ in $AdS_{4}$. In free theory limit, it is 
\begin{equation}
h^{\mu_{1}\cdots\mu_{s}}=	W^{\{\mu_{1}}_{a_{1}}W^{\mu_{s-1}}_{a_{s-1}}W^{\mu_{s}\}}_{a_{s}}g^{a_{s}a}W^{\mu}_{a}W^{a_{1}\cdots a_{s-1},0\cdots 0}_{\mu}
\end{equation}
that will finally appear in equations of motion and the gauge transformation. One may expect in interacting case, the final dynamics is also expressed in terms of some $h^{\mu_{1}\cdots\mu_{s}}	$, which can be a more complicated combination of $W^{a_{1}\cdots a_{s-1},0\cdots 0}_{\mu}$. The frame-like formulation reduces to the metric-like formulation.

Altogether, the complete equations are 
\begin{eqnarray}
 \label{1g} && 	\hat{f}^{\alpha}_{\beta\gamma}=\hat{f}^{\alpha}_{\beta\gamma} (R^{[a(t-1),b(t-1)]}_{ab},R^{[a(t-1),b(t-1)]}_{ab;c_{1}},\cdots,H,H_{c_{1}},\cdots), \\  & &
  	H_{\gamma}=h_{\gamma} (R^{[a(t-1),b(t-1)]}_{ab},R^{[a(t-1),b(t-1)]}_{ab;c_{1}},\cdots,H,H_{c_{1}},\cdots)
  ,\label{2g} \\ & &
r^{[a(s-1),b(s-1)]}_{ab;c_{1}\cdots c_{n} \gamma} = r^{[a(s-1),b(s-1)]}_{ab;c_{1}\cdots c_{n} \gamma}(R^{[a(t-1),b(t-1)]}_{ab},R^{[a(t-1),b(t-1)]}_{ab;c_{1}},\cdots,H,H_{c_{1}},\cdots),\label{3g} \\  & &
h_{c_{1}\cdots c_{n} \gamma} =h_{c_{1}\cdots c_{n} \gamma} (R^{[a(t-1),b(t-1)]}_{ab},R^{[a(t-1),b(t-1)]}_{ab;c_{1}},\cdots,H,H_{c_{1}},\cdots), \label{4g}  
\end{eqnarray}
\begin{eqnarray}
& & d W^{\alpha} =\frac{1}{2}\hat{f}^{\alpha}_{\beta\gamma}W^{\beta}\wedge W^{\gamma}, \label{5g}  \\  & & d R^{[a(s-1),b(s-1)]}_{ab;c_{1}\cdots c_{n}} = r^{[a(s-1),b(s-1)]}_{ab;c_{1}\cdots c_{n} \gamma}W^{\gamma}\Leftrightarrow \partial_{\gamma} R^{[a(s-1),b(s-1)]}_{ab;c_{1}\cdots c_{n}} = r^{[a(s-1),b(s-1)]}_{ab;c_{1}\cdots c_{n} \gamma},\label{6g}\\  & & d H_{c_{1}\cdots c_{n}} = h_{c_{1}\cdots c_{n} \gamma}W^{\gamma}\Leftrightarrow \partial_{\gamma} H_{c_{1}\cdots c_{n}} = h_{c_{1}\cdots c_{n} \gamma},\label{7g}  
\end{eqnarray}
\begin{eqnarray}
 & &  r^{[a(s-1),b(s-1)]}_{ab;c_{1} \cdots c_{n}[\gamma}\frac{\partial \hat{f}^{\alpha}_{\rho\sigma]}}{\partial R^{[a(s-1),b(s-1)]}_{ab;c_{1} \cdots c_{n}}} +h_{c_{1} \cdots c_{n}[\gamma}\frac{\partial \hat{f}^{\alpha}_{\rho\sigma]}}{\partial H_{c_{1} \cdots c_{n}}}+\hat{f}^{\alpha}_{\beta[\gamma}\hat{f}^{\beta}_{\rho\sigma]}=0,\label{8g}  \\  & &   r^{[a(s-1),b(s-1)]}_{ab;c_{1} \cdots c_{n}[\rho}\frac{\partial h_{\sigma]}}{\partial R^{[a(s-1),b(s-1)]}_{ab;c_{1} \cdots c_{n}}} +h_{c_{1} \cdots c_{n}[\rho}\frac{\partial h_{\sigma]}}{\partial H_{c_{1} \cdots c_{n}}}+h_{\alpha}\hat{f}^{\alpha}_{\rho\sigma}=0.\label{9g}
\end{eqnarray}
The input is $(\hat{f}^{\alpha}_{\beta\gamma}, h_{\gamma})$, from which, all the rest equations are determined. The left hand sides of the $4d$ equations of motion (\ref{8g})-(\ref{9g}) are polynomials of $(R^{[a(s-1),b(s-1)]}_{ab;c_{1} \cdots c_{n}}, H_{c_{1} \cdots c_{n}})$. For the randomly selected $(\hat{f}^{\alpha}_{\beta\gamma}, h_{\gamma})$, (\ref{8g})-(\ref{9g}) only has the trivial solution $R^{[a(s-1),b(s-1)]}_{ab;c_{1} \cdots c_{n}}= H_{c_{1} \cdots c_{n}}=0$. A natural question is what might be the maximum on-shell degrees of freedom. If one can find such $(\hat{f}^{\alpha}_{\beta\gamma}, h_{\gamma})$, for which, (\ref{8g})-(\ref{9g}) is satisfied for the arbitrary $(R^{[a(s-1),b(s-1)]}_{ab;c_{1} \cdots c_{n}}, H_{c_{1} \cdots c_{n}})$, then there are no $4d$ equations of motion. This is not quite likely to be the case.\footnote{If it is true, then the $4d$ local HS gauge transformation (\ref{GAU}) can be closed off-shell (for the arbitrary $W_{\mu}^{[a(s-1),b(0)]}$ and $H$ in $AdS_{4}$).} By partially solving (\ref{8g})-(\ref{9g}), one may determine $(\hat{f}^{\alpha}_{\beta\gamma}, h_{\gamma})$, which, when plugged in (\ref{8g})-(\ref{9g}), gives the $4d$ equations of motion for $(R^{[a(s-1),b(s-1)]}_{ab;c_{1} \cdots c_{n}}, H_{c_{1} \cdots c_{n}})$. In supergravity situation, the procedure is quite simple as is demonstrated in Section 3.2. In higher spin theory, the more efficient way is to first determine the on-shell degrees of freedom $\Phi^{\tilde{\alpha}}$. Then with the off-shell $(R^{[a(s-1),b(s-1)]}_{ab;c_{1} \cdots c_{n}}, H_{c_{1} \cdots c_{n}})$ expressed in terms of the on-shell $\Phi^{\tilde{\alpha}}$, we only need to find $(\hat{f}^{\alpha}_{\beta\gamma}, h_{\gamma})$ satisfying the Bianchi identity for the arbitrary $\Phi^{\tilde{\alpha}}$. From the on-shell $(R^{[a(s-1),b(s-1)]}_{ab;c_{1} \cdots c_{n}}, H_{c_{1} \cdots c_{n}})$ at one point, or the on-shell $(W^{[a(s-1),b(0)]}_{\mu},H)$ in $AdS_{4}$, $(W^{\alpha},H)$ on $\textbf{M}$ can be determined via (\ref{5g})-(\ref{7g}). With $(W^{\alpha},H)$ on $\textbf{M}$ solved, the finite local higher spin transformation is the finite diffeomorphism transformation on $\textbf{M}$, under which, $(W^{[a(s-1),b(0)]}_{\mu},H)$ in one $AdS_{4}$ slice is moved to $(W^{[a(s-1),b(0)]}_{\mu},H)$ in another $AdS_{4}$ slice. The higher spin symmetry is realized as an on-shell symmetry.

\subsection{Relation with the unfolded equation in Vasiliev theory}

With $\Phi^{\tilde{\alpha}}$ representing the on-shell degrees of freedom of $(R^{[a(s-1),b(s-1)]}_{ab;c_{1} \cdots c_{n}}, H_{c_{1} \cdots c_{n}})$, where $\tilde{\alpha}$ is in some representation of the Lorentz group, the unfolded equation becomes 
\begin{eqnarray}
 && 	\bar{f}^{\alpha}_{\beta\gamma}=\bar{f}^{\alpha}_{\beta\gamma} (\Phi^{\tilde{\sigma}}), \;\;\;\;\;\;\;\;\;\; \;\;\;\;\;\;\;\;\;\; \;\;\;
F^{\tilde{\alpha}}_{ \gamma} = F^{\tilde{\alpha}}_{ \gamma}(\Phi^{\tilde{\sigma}}),  \label{3w1}
   \\ & & d W^{\alpha} =\frac{1}{2}\bar{f}^{\alpha}_{\beta\gamma}W^{\beta}\wedge W^{\gamma},  \;\;\;\;\;\;\;\;\;\;  d \Phi^{\tilde{\alpha}}= F^{\tilde{\alpha}}_{\beta} W^{\beta} , \label{3w11} \\  & &  F^{\tilde{\beta}}_{ [\gamma}	\frac{\partial \bar{f}^{\alpha}_{\rho\sigma]}}{\partial \Phi^{\tilde{\beta}}}+\bar{f}^{\alpha}_{\beta[\gamma}\bar{f}^{\beta}_{\rho\sigma]}=0 , \;\;\;\;\;\;\;\;\;\;  \frac{\partial F^{\tilde{\alpha}}_{[\sigma}}{\partial \Phi^{\tilde{\beta}}}F^{\tilde{\beta}}_{\rho]}+F^{\tilde{\alpha}}_{\gamma}\bar{f}^{\gamma}_{\rho\sigma}  =0       .\label{6gy7}
\end{eqnarray}
It remains to find the suitable $(\bar{f}^{\alpha}_{\beta\gamma},F^{\tilde{\alpha}}_{ \gamma} )$ with the Bianchi identity (\ref{6gy7}) satisfied for the arbitrary $\Phi^{\tilde{\alpha}}$. Under the field redefinition $\Phi^{\tilde{\alpha}}\rightarrow \varphi^{\tilde{\alpha}}=\varphi^{\tilde{\alpha}}(\Phi^{\tilde{\sigma}})$, $F^{\tilde{\alpha}}_{\beta}\rightarrow \frac{\partial \varphi^{\tilde{\alpha}}}{\partial \Phi^{\tilde{\sigma}}}F^{\tilde{\sigma}}_{\beta}$.

In Vasiliev theory, $\Phi^{\tilde{\alpha}}\sim \Phi^{[a(s+n),b(s)]}$ is in the twisted-adjoint representation of the higher spin algebra. (\ref{3w1}) is obtained by solving the Vasiliev equation order by order. (\ref{6gy7}) is then automatically satisfied for the arbitrary $\Phi^{\tilde{\alpha}}$. Let us make a comparison between
\begin{equation}\label{dcgy5}
	\{H_{c_{1} \cdots c_{n}}|n=0,1,\cdots\}\;\cup \;\{R^{[a(s-1),b(s-1)]}_{ab;c_{1} \cdots c_{n}}| s=2,4,\cdots, n \\ =0,1,\cdots\}
\end{equation}
and $\{\Phi^{[a(s+n),b(s)]}, s=0,2,\cdots, n=0,1,\cdots\}$. The two have the same number of indices, but the former is the off-shell field while the latter is on-shell. With the $4d$ equations of motion imposed on (\ref{dcgy5}), the two may contain the same number of degrees of freedom.

Fields in the twisted-adjoint representation and the adjoint representation are related via the action of the Klein operator \cite{mi}
\begin{equation}
	\Phi^{\alpha}t_{\alpha}\rightarrow \Phi^{\alpha}t_{\alpha}\ast \kappa= \Phi^{\alpha}\rho^{\tilde{\alpha}}_{\alpha}t_{\tilde{\alpha}} =\Phi^{\tilde{\alpha}}t_{\tilde{\alpha}}~,
\end{equation}
where $\Phi^{\tilde{\alpha}} = \rho^{\tilde{\alpha}}_{\alpha}\Phi^{\alpha}$, $\Phi^{\alpha} = (\rho^{-1})^{\alpha}_{\tilde{\alpha}}\Phi^{\tilde{\alpha}}$, $\rho^{\tilde{\alpha}}_{\alpha}$ is a constant matrix. For $\Phi$ in adjoint representation, i.e. $\Phi^{\alpha} \sim \Phi^{[a(s-1),b(t)]}$, it is possible to let $F^{\alpha}_{\beta}=\bar{f}^{\alpha}_{\beta\gamma}\Phi^{\gamma}$ \cite{3h}, (\ref{3w1})-(\ref{6gy7}) reduces to
\begin{eqnarray}
\label{g1}  && 	\bar{f}^{\alpha}_{\beta\gamma}=\bar{f}^{\alpha}_{\beta\gamma} (\Phi^{\sigma}),
   \\  & & d W^{\alpha} =\frac{1}{2}\bar{f}^{\alpha}_{\beta\gamma}W^{\beta}\wedge W^{\gamma}, \;\;\;\;\;\;\;\;\;\;  d \Phi^{\alpha}=\bar{f}^{\alpha}_{\beta\gamma} \Phi^{\gamma} W^{\beta}, \label{g2}  \\  & &  -\Phi^{\nu}\bar{f}^{\beta}_{\nu [\gamma}	\frac{\partial \bar{f}^{\alpha}_{\rho\sigma]}}{\partial \Phi^{\beta}}+\bar{f}^{\alpha}_{\beta[\gamma}\bar{f}^{\beta}_{\rho\sigma]}=0       .\label{g3} 
\end{eqnarray}
With $\Phi^{\alpha}\rightarrow \Phi^{\tilde{\alpha}}$, (\ref{3w1})-(\ref{6gy7}) is recovered for $F^{\tilde{\alpha}}_{\beta}=\bar{k}^{\tilde{\alpha}}_{\beta\tilde{\gamma}} \Phi^{\tilde{\gamma}}$. $\bar{k}^{\tilde{\alpha}}_{\beta\tilde{\gamma}} =\rho^{\tilde{\alpha}}_{\alpha}\rho^{\gamma}_{\tilde{\gamma}} \bar{f}^{\alpha}_{\beta\gamma}$. $\Phi^{[a(0),b(0)]}\equiv\Phi=H$, $\partial_{\beta}\Phi = \partial_{\beta} H = H_{\beta}=\bar{k}_{\beta\tilde{\gamma}}\Phi^{\tilde{\gamma}}$. $\partial_{\beta}\Phi^{\tilde{\alpha}} = \bar{k}^{\tilde{\alpha}}_{\beta\tilde{\gamma}} \Phi^{\tilde{\gamma}}$.

With (\ref{3w1})-(\ref{6gy7}) at hand, we have $R^{\alpha}_{\beta\gamma}=R^{\alpha}_{\beta\gamma}(\Phi^{\tilde{\sigma}})=\bar{f}^{\alpha}_{\beta\gamma}(\Phi^{\tilde{\sigma}})-f^{\alpha}_{\beta\gamma}$, $H_{\beta}=F_{\beta}(\Phi^{\tilde{\sigma}})$ since $d\Phi=F_{\beta}W^{\beta}$. Especially, 
\begin{equation}\label{1988}
R^{[a(s-1),b(s-1)]}_{ab}=	R^{[a(s-1),b(s-1)]}_{ab}(\Phi^{\tilde{\sigma}})  ,\;\;\;\;\;\;\;\;\;\;H=\Phi,
\end{equation}
and subsequently, 
\begin{equation}\label{2046}
R^{[a(s-1),b(s-1)]}_{ab;c_{1}\cdots c_{n}}=	R^{[a(s-1),b(s-1)]}_{ab;c_{1}\cdots c_{n}}(\Phi^{\tilde{\sigma}})  ,\;\;\;\;\;\;\;\;\;	H_{c_{1}\cdots c_{n}} = H_{c_{1}\cdots c_{n}} (\Phi^{\tilde{\sigma}}),
\end{equation}
where $\partial_{c} \Phi^{\tilde{\alpha}} = F^{\tilde{\alpha}}_{c}(\Phi^{\tilde{\sigma}})$ is used.
\begin{eqnarray}\label{g3d}
\nonumber   && 	\hat{f}^{\alpha}_{\beta\gamma}(R^{[a(s-1),b(s-1)]}_{ab;c_{1}\cdots c_{n}},H_{c_{1}\cdots c_{n}})=\hat{f}^{\alpha}_{\beta\gamma}[R^{[a(s-1),b(s-1)]}_{ab;c_{1}\cdots c_{n}}(\Phi^{\tilde{\sigma}}),H_{c_{1}\cdots c_{n}}(\Phi^{\tilde{\sigma}})]=\bar{f}^{\alpha}_{\beta\gamma}(\Phi^{\tilde{\sigma}})
    \\  & & h_{\gamma}(R^{[a(s-1),b(s-1)]}_{ab;c_{1}\cdots c_{n}},H_{c_{1}\cdots c_{n}})=h_{\gamma}[R^{[a(s-1),b(s-1)]}_{ab;c_{1}\cdots c_{n}}(\Phi^{\tilde{\sigma}}),H_{c_{1}\cdots c_{n}}(\Phi^{\tilde{\sigma}})]=F_{\gamma}(\Phi^{\tilde{\sigma}})  
\end{eqnarray}

Let us return to the discussion below (\ref{9g}). With $\hat{f}^{\alpha}_{\beta\gamma}$ and $h_{\gamma}$ determined by the Bianchi identity, (\ref{8g})-(\ref{9g}) may still have further constraints on $(R^{[a(s-1),b(s-1)]}_{ab;c_{1}\cdots c_{n}},H_{c_{1}\cdots c_{n}})$, which are the $4d$ equations of motion. Alternatively, one may use the on-shell $\Phi^{\tilde{\sigma}}$ to parameterize the off-shell $(R^{[a(s-1),b(s-1)]}_{ab;c_{1}\cdots c_{n}},H_{c_{1}\cdots c_{n}})$ as is in (\ref{2046}). $4d$ equations of motion are then solved automatically. (\ref{8g})-(\ref{9g}) does not impose any constraints on $\Phi^{\tilde{\sigma}}$. The key step in group manifold approach is to get the rheonomy condition and the $4d$ equations of motion from the Bianchi identity. For higher spin theory, the on-shell degrees of freedom form the twisted-adjoint representation of the higher spin algebra, while the Vasiliev equation gives an elegant way to solve the Bianchi identity. The solution for $(W^{\alpha},H)$ in $\textbf{M}$ is characterized by the on-shell $(R^{[a(s-1),b(s-1)]}_{ab;c_{1}\cdots c_{n}},H_{c_{1}\cdots c_{n}})$ at one point, or by the arbitrary $\Phi^{\tilde{\alpha}}$ at that point. Nevertheless, it is $(R^{[a(s-1),b(s-1)]}_{ab;c_{1}\cdots c_{n}},H_{c_{1}\cdots c_{n}})$ that has the physical meaning. We are free to make a change of the variables $\varphi^{\tilde{\alpha}}=\varphi^{\tilde{\alpha}}(\Phi^{\tilde{\sigma}})$ to use $\varphi^{\tilde{\alpha}}$ to parameterize $(R^{[a(s-1),b(s-1)]}_{ab;c_{1}\cdots c_{n}},H_{c_{1}\cdots c_{n}})$. The good variables are those which are as relevant to $(R^{[a(s-1),b(s-1)]}_{ab;c_{1}\cdots c_{n}},H_{c_{1}\cdots c_{n}})$ as possible.

The nonlinear higher spin theory should also have the proper free theory limit that is equivalent to Fronsdal theory \cite{1, 68}. In free theory limit, the equations of motion in (\ref{1g})-(\ref{9g}) become
\begin{eqnarray}
  \label{236}  && 	d W^{\alpha} - f^{\alpha}_{(ab)\gamma}W^{(ab)}\wedge W^{\gamma}- f^{\alpha}_{a \gamma}W^{a}\wedge W^{\gamma}=\frac{1}{2}R^{\alpha}_{ab}W^{a}\wedge W^{b},
   \\  \label{321s} & & 	d R^{[a(s-1),b(s-1)]}_{ab;c_{1}\cdots c_{n}} =r^{[a(s-1),b(s-1)]}_{ab;c_{1}\cdots c_{n}(cd)}W^{(cd)}+R^{[a(s-1),b(s-1)]}_{ab;c_{1}\cdots c_{n}c}W^{c},  \\ \label{321ss} & &  d H_{c_{1}\cdots c_{n}}=h_{c_{1}\cdots c_{n}(cd)}W^{(cd)}+H_{c_{1}\cdots c_{n}c_{n+1}}W^{c_{n+1}}.
\end{eqnarray}
Since $r^{[a(s-1),b(s-1)]}_{ab;c_{1}\cdots c_{n}(cd)} = \partial_{(cd)}R^{[a(s-1),b(s-1)]}_{ab;c_{1}\cdots c_{n}}$ and $h_{c_{1}\cdots c_{n}(cd)} = \partial_{(cd)}H_{c_{1}\cdots c_{n}}$ give the local Lorentz transformation, (\ref{321s})-(\ref{321ss}) can be rewritten as
\begin{eqnarray}
    && 		D R^{[a(s-1),b(s-1)]}_{ab;c_{1}\cdots c_{n}} =R^{[a(s-1),b(s-1)]}_{ab;c_{1}\cdots c_{n}c_{n+1}}W^{c_{n+1}},  \\ & &  	D H_{c_{1}\cdots c_{n}}=H_{c_{1}\cdots c_{n}c_{n+1}}W^{c_{n+1}}~,
\end{eqnarray}
where $D$ is the standard covariant derivative with the connection $W^{(cd)}$. $DH=dH =H_{c}W^{c} $. For the theory to have the correct free theory limit, there will be 
\begin{equation}\label{4587}
	R^{[a(s-1),b(t)]}_{ab}=0 \;\;\;\;\;\;\;\;\; \textnormal{for} \;\;\;\;\;\;\;\;\; t \neq s-1
\end{equation}
so that (\ref{236}) becomes 
\begin{eqnarray}\label{387}
  \nonumber  && 			D W^{[a(s-1),b(t)]} = f^{[a(s-1),b(t)]}_{a [c(s-1),d(t+1)]}W^{a}\wedge W^{[c(s-1),d(t+1)]}+ f^{[a(s-1),b(t)]}_{a [c(s-1),d(t-1)]}W^{a}\wedge W^{[c(s-1),d(t-1)]} ~,\\ \nonumber & &  	D W^{[a(s-1),b(s-1)]}= f^{[a(s-1),b(s-1)]}_{a [c(s-1),d(s-2)]}W^{a}\wedge W^{[c(s-1),d(s-2)]}+\frac{1}{2}R^{[a(s-1),b(s-1)]}_{ab}W^{a}\wedge W^{b}~,\\
\end{eqnarray}
where $t<s-1$. (\ref{4587}) is also called the ``central on-mass-shell theorem''\cite{cen, cen1}. In Vasiliev theory, $R^{\alpha}_{\beta\gamma}$ satisfies (\ref{4587}) at the first order of the $\Phi^{\tilde{\alpha}}$ expansion.

Since the adjoint representation and the twisted-adjoint representation are related by a Klein transformation which is invertible, we may try to use $\Phi^{\alpha}$ to parameterize $\bar{f}^{\alpha}_{\beta\gamma}$ as is in (\ref{g1}). If we further make a restriction that (\ref{g2}) can be written as 
\begin{equation}
	d W= H(W,\Phi),\;\;\;\;\;\;\;\;\;\;\;\;\;\;d \Phi = F(W, \Phi)
\end{equation}
with $H(W,\Phi)$ and $F(W, \Phi)$ polynomials of $W=W^{\alpha}t_{\alpha}$ and $\Phi = \Phi^{\alpha}t_{\alpha}$, the solution for (\ref{g3}) can be easy fixed, which is given in Appendix C. Although the Bianchi identity is satisfied for the arbitrary $\Phi^{\alpha}$, (\ref{4587}) does not hold at the first order of the $\Phi^{\alpha}$ expansion, so the theory does not have the correct free theory limit.

Satisfying the Bianchi identity for the on-shell $(R^{[a(s-1),b(s-1)]}_{ab;c_{1}\cdots c_{n}},H_{c_{1}\cdots c_{n}})$ and giving rise to the correct free theory limit are two requirements for $(\hat{f}^{\alpha}_{\beta\gamma},h_{\gamma})$. It is unclear whether the requirements can uniquely fix $(\hat{f}^{\alpha}_{\beta\gamma},h_{\gamma})$ or not. Starting from the the rheonomy condition (\ref{4r67h}) in Section 3.1, one may get (\ref{gvxfg1}) with no torsion constraint imposed on $W^{\alpha}_{\mu}$. The torsion constraint is just (\ref{4587}), or concretely, 
\begin{equation}
	R^{[a(s-1),b(t)]}_{ab} = g^{[a(s-1),b(t)]}_{ab}(W^{\sigma}_{\mu},\partial_{\nu_{1}}W^{\sigma}_{\mu},\cdots, H,\partial_{\nu_{1}}H,\cdots )=0,\;\; \textnormal{for} \;\; t \neq s-1,
\end{equation}
which will make $W^{\alpha}_{\mu}$ reduce to $W^{[a(s-1),b(0)]}_{\mu}$ and also guarantee the correct free theory limit. In this case, having the correct free theory limit and satisfying the torsion constraint are the same thing. If there is such (\ref{4r67h}), for which the Bianchi identity on $(R^{\alpha}_{ab;c_{1}\cdots c_{n}},H_{c_{1}\cdots c_{n}})$ reduces to the $4d$ equations of motion, then by setting $R^{[a(s-1),b(t)]}_{ab}$ to $0$ for $ t \neq s-1$, we will get (\ref{vgyh}) satisfying the Bianchi identity for the on-shell $(R^{[a(s-1),b(s-1)]}_{ab;c_{1}\cdots c_{n}},H_{c_{1}\cdots c_{n}})$ and having the correct free theory limit. (\ref{4587}) holds exactly in this situation.

Finally, we will have a heuristic discussion on the group manifold approach to conformal HS theory. $3d$ conformal HS algebra and $AdS_{4}$ HS algebra are the same, so the corresponding group manifold is also $\textbf{M}$. The equations are still
\begin{equation}\label{cccs}
	dW^{\alpha}=\frac{1}{2}(f^{\alpha}_{\beta\gamma}+R^{\alpha}_{\beta\gamma})W^{\beta}\wedge W^{\gamma},\;\;\;\;\;\;\;dH=H_{\alpha}W^{\alpha}.
\end{equation}
The submanifold of interest is not $AdS_{4}$ but $\partial AdS_{4} \subset \partial\textbf{M}$. The solution of the unfolded equation in $\textbf{M}$ is determined by the value of the 0-form multiplet at one point. In previous discussion, this point is selected at the bulk of $AdS$, but now, it should live at $\partial AdS$. The generated solution will remain at the near boundary region, since an infinite evolution is needed to move from the boundary to the bulk. The rheonomy condition $\bar{f}^{\alpha}_{\beta\gamma}=\bar{f}^{\alpha}_{\beta\gamma}(\Phi^{\tilde{\sigma}})$ and $F^{\tilde{\alpha}}_{ \gamma} = F^{\tilde{\alpha}}_{ \gamma}(\Phi^{\tilde{\sigma}})$ in 
Vasiliev theory may undergo a reduction at the boundary with the role of $\Phi^{\tilde{\sigma}}$ played by a smaller set of 0-forms so that the solution at the near boundary region is determined by the dynamical fields in $3d$.

In $ho(1|2:[3,2])$, the dilaton is $t_{0,4}=D$. It is convenient to choose the basis $\{t_{\alpha}\}$ with the definite conformal dimension, i.e. $[D,t_{\alpha}]=i\Delta_{\alpha}t_{\alpha}$. For example, the basis of $so(3,2)$ is $\{D,P_{i},K_{i},L_{i,j}\}$ with $i,j=1,2,3$. The dynamical fields are $W^{i_{1}\cdots i_{s-1}}_{m}$ in $3d$, $m=1,2,3$ \cite{aqq2}. Here $i_{1}\cdots i_{s-1}$ refers to the index of the spin $s$ generator $P_{i_{1}\cdots i_{s-1}}$ with the dimension $1-s$.

There is a conjecture that the conformal HS theory at $\partial AdS_{d+1}$ is related to the HS theory in $AdS_{d+1}$ with the action of the conformal HS fields for even $d$ equals to the logarithmically divergent term of the action of HS fields in $AdS_{d+1}$ \cite{aqq222, aqq22}. In \cite{vas}, the unfolded equation for a $3d$ conformal HS theory coming from the boundary limit of the $AdS_{4}$ Vasiliev theory was considered. It was shown that at $\partial AdS_{4}$, $R^{\alpha}_{ab}\neq 0$ only when $t_{\alpha}=K_{i_{1}\cdots i_{s-1}}$. The condition could make $W^{\alpha}_{m}$ expressed in terms of the dynamical field $W^{i_{1}\cdots i_{s-1}}_{m}$ without imposing constraints on the latter. Correspondingly, in (\ref{vgyh}), the independent 0-forms are $(R^{\alpha}_{ij},R^{\alpha}_{ij;k_{1}},\cdots,H,H_{k_{1}},\cdots)$ for $t_{\alpha}=K_{i_{1}\cdots i_{s-1}}$, $s=2,4,\cdots$. This is consistent with the fact that in odd dimensions, the conformal HS theory is trivial with no equations of motion imposed on dynamical fields \cite{aqq2, aqq22, aqq3}.

On the other hand, in even dimensions, dynamical fields should satisfy Fradkin-Tseytlin equation \cite{confk}. The unfolded system of Fradkin-Tseytlin equation was formulated in \cite{confk1, confk2}, where the 0-form multiplet is Weyl module generated by Weyl tensor, which, according to the terminology of \cite{aqq2}, is the ground field strength. Equivalently, the 0-forms in (\ref{vgyh}) should now be taken as $(R^{[i(s-1),j(s-1)]}_{ij},R^{[i(s-1),j(s-1)]}_{ij;k_{1}},\cdots,H,H_{k_{1}},\cdots)$. In free theory limit, $R^{\alpha}_{ij}=0$ if $\Delta_{\alpha}<0$, $R^{\alpha}_{ij}$ with $\Delta_{\alpha}>0$ can all be expressed in terms of the derivatives of the Weyl tensor $R^{[i(s-1),j(s-1)]}_{ij}$. This is somewhat different from \cite{vas} for $3d$, where $R^{[i(s-1),j(s-1)]}_{ij}=0$. It is interesting to consider the $4d$ conformal HS system arising from the boundary reduction of the Vasiliev equation in $ AdS_{5}$ in analogy with \cite{vas}. In free theory limit, the obtained equation is expected to give the unfolded system of Fradkin-Tseytlin equation \cite{confk1, confk2}. The boundary value of the $AdS_{d+1}$ HS fields was considered in \cite{aqq3, aqq4} in the ambient approach, where it was shown that for even $d$ there is an obstruction for the bulk extension unless the conformal HS fields at $\partial AdS_{d+1}$ satisfy the Fradkin-Tseytlin equation. In this case, the near boundary expansion of the on-shell $AdS$ field (see, for examaple, \cite{4r}) does not have the logarithm term, which is required in the unfolded formalism, which in the minimal version does not allow for logarithmic terms to cancel the obstruction.

The unfolded equation for the $4d$ HS theory is invariant under the local Lorentz transformation $SO(3,1)$, i.e. $R^{\alpha}_{(a,b)\gamma}=0$. In \cite{vas}, it is possible to impose the suitable boundary condition so that $R^{\alpha}_{(0,4)i}=0$. If the conclusion can be extended to $R^{\alpha}_{(0,4)\gamma}=0$, then the dilatation is unformed. Moreover, the original HS theory already have the undeformed $SO(3,1)$ local Lorentz transformation, so by a naive counting, it seems that the inhomogeneous Weyl group $\mathcal{IW}$ generated by $\{D,K_{i},L_{i,j}\}$ can be undeformed at the boundary. In this case, the evolution along the $t_{0,4}$ direction is a conformal (gauge) transformation and the dynamics is reduced from $4d$ to $3d$. It remains to see whether there are consistent nonlinear unfolded equation for the conformal HS theory meeting this requirement. At least, the $3d$ local Lorentz transformation is undeformed.

\subsection{The extended action principle for higher spin theory}

In group manifold approach to supergravity, instead of imposing the rheonomy condition directly, one may construct the extended action whose variation gives both the rheonomy condition and the $4d$ equations of motion \cite{1g}.

For example, in $\mathcal{N}=1$ supergravity, the extended action is of the form 
\begin{equation}
	S=S[\nu^{A},M_{4}]=\int_{M_{4} \subset M} L^{(4)} (\nu^{A})~,
\end{equation}
where $M_{4}$ is a $4d$ submanifold of the superspace $M$,\footnote{We can use the group manifold $\textbf{M}$ instead of $M$, but the result is the same due to the factorization condition.} and $L^{(4)}$ is a local Lorentz invariant $4$-form in $M$ constructed from $\nu^{A}$ via the exterior differentiation and the exterior product. Variation of $S$ with respect to both $\nu^{A}$ and $M_{4}$ gives
\begin{equation}
	\frac{\delta L^{(4)}}{\delta \nu^{A}} = K^{(3)}_{A}(z)=0.
\end{equation}
$K^{(3)}_{A}$ is a 3-form that should vanish all over $M$. $ K^{(3)}_{A}(z)=0$ contains both the rheonomy condition and the $4d$ equations of motion. The concrete form of $L^{(4)} $ is  
\begin{equation}
	L^{(4)} =\epsilon_{abcd}R^{ab} \wedge \nu^{c}\wedge \nu^{d} + 4 \bar{\psi} \wedge \gamma_{5}\gamma_{a} \rho \wedge \nu^{a},
\end{equation}
where $\rho^{\alpha}_{MN}=R^{\alpha}_{MN}$.

For higher spin theory, if the extended action exists, it takes the form
\begin{equation}
	S=S[W^{\alpha},M_{4}]=\int_{M_{4} \subset \textbf{M}} L^{(4)}(W^{\alpha}), 
\end{equation}
where $L^{(4)}$ is a $4$-form invariant under the local Lorentz transformation. 
\begin{equation}
	K^{(3)}_{\sigma}=\frac{\delta L^{(4)}}{\delta W^{\sigma}} =K^{(3)}_{\sigma[\alpha\beta\gamma]}W^{\alpha}\wedge W^{\beta} \wedge W^{\gamma},
\end{equation}
\begin{equation}
			K^{(3)}_{\sigma}=0 \Leftrightarrow K^{(3)}_{\sigma[\alpha\beta\gamma]}=0.
\end{equation}
We need to find the configuration $W^{\alpha}$ on $\textbf{M}$ with $K^{(3)}_{\sigma}=0 $ everywhere. Still, the on-shell solution on $\textbf{M}$ is characterized by the on-shell solution on $M_{4}$. $M_{4} \rightarrow M'_{4}$ is a diffeomorphism transformation on $\textbf{M}$ that is equivalent to the deformed higher spin gauge transformation. The equation $K^{(3)}_{\sigma}=0$ is on-shell gauge invariant. Off-shell higher spin invariance has the further requirement $d L^{(4)}=0$ \cite{1g}. Although the on-shell gauge invariance is automatically guaranteed, for the generic $L^{(4)}$, $K^{(3)}_{\sigma}=0$ only has the trivial solution $R^{\alpha}_{\beta\gamma}=0$, so the question is whether there is $L^{(4)}$ for which, the related $K^{(3)}_{\sigma}=0$ has the nontrivial solution or not. In supergravity, having the nontrivial solution also puts the severe constraint on $S$.

In the simplest situation, if 
\begin{equation}
	L^{(4)}= \kappa_{\alpha\beta}R^{\alpha}\wedge R^{\beta}+\kappa_{\alpha\beta\gamma}R^{\alpha}\wedge W^{\beta}\wedge W^{\gamma}+\kappa_{\alpha\beta\gamma\sigma}W^{\alpha}\wedge W^{\beta}\wedge W^{\gamma}\wedge W^{\sigma}
\end{equation}
with $\kappa$ constants, then 
\begin{equation}\label{im}
K^{(3)}_{\sigma[\alpha\beta\gamma]}=-2\kappa_{\sigma\rho}f^{\rho}_{\chi[\alpha}R^{\chi}_{\beta\gamma]}-2\kappa_{\rho\chi}f^{\rho}_{\sigma[\alpha}R^{\chi}_{\beta\gamma]}+\kappa_{\sigma\rho[\gamma}\hat{f}^{\rho}_{\alpha\beta]}+2\kappa_{\rho\sigma[\gamma}R^{\rho}_{\alpha\beta]}+\kappa_{\rho[\beta\gamma}f^{\rho}_{\alpha]\sigma}+4\kappa_{\sigma[\alpha\beta\gamma]}	.
\end{equation}
(\ref{im}) imposes a set of linear relations among $R^{\alpha}_{\beta\gamma}$, which, when plugged into the Bianchi identity, may only allow the trivial solution $R^{\alpha}_{\beta\gamma}=0$. The more general form of $L^{(4)}$ is 
\begin{equation}\label{lala}
L^{(4)}=f_{\rho\sigma\chi\eta}(R^{\alpha}_{\beta\gamma},\partial_{\lambda}R^{\alpha}_{\beta\gamma},\cdots)	W^{\rho}\wedge W^{\sigma}\wedge W^{\chi}\wedge W^{\eta}
\end{equation}
including an infinite number of derivatives. $K^{(3)}_{\sigma[\alpha\beta\gamma]}=0$ are functions of $(R^{\alpha}_{\beta\gamma},\partial_{\lambda}R^{\alpha}_{\beta\gamma},\cdots)$. With $R^{\alpha}_{\beta\gamma} = \bar{f}^{\alpha}_{\beta\gamma}(\Phi^{\tilde{\sigma}})-f^{\alpha}_{\beta\gamma}$ plugged in, $K^{(3)}_{\sigma[\alpha\beta\gamma]}$ should automatically vanish for the arbitrary $\Phi^{\tilde{\sigma}}$ if it is the action from which, the Vasiliev equation comes out. However, it is too complicated to fix the exact form of (\ref{lala}).

\subsection{Dynamics of 0-form matter on group manifold with the fixed background }

(\ref{1g})-(\ref{9g}) describes the coupling of the spin $0$ matter $H$ and the spin $2, 4, \cdots$ gravity field $W^{\alpha}$. Under the local gauge transformation, which is the deformed higher spin transformation as well as the diffeomorphism transformation on $\textbf{M}$, spin $0, 2, 4, \cdots$ fields mix with each other. To describe the dynamics of the $0$-form matter on $\textbf{M}$ with the fixed background, the matter-gravity coupling must be turned off. One may let $r^{\alpha}_{\beta\gamma}=0$, then $W_{0}^{\alpha}$ gives the intrinsic geometry of the group manifold $\textbf{M}$ discussed in Section 2. The equations of motion reduce to 
\begin{equation}\label{204}
	d W_{0}^{\alpha} -\frac{1}{2}f^{\alpha}_{\beta\gamma}W_{0}^{\beta}\wedge W_{0}^{\gamma} =0,\;\;\;\;\;\;\;d H = H_{\alpha} W_{0}^{\alpha}\Leftrightarrow \partial_{\alpha}H=H_{\alpha},\;\;\;\;\;\;\;\partial_{[\rho}H_{\sigma]}+H_{\alpha}f^{\alpha}_{\rho\sigma} = 0.
\end{equation}
The allowed gauge transformation parameter $\epsilon^{\alpha}_{0}$ should satisfy
\begin{equation}\label{fc35}
	\delta_{\epsilon_{0}} W_{0\; \bar{M}}^{\alpha}=\partial_{\bar{M}} \epsilon_{0}^{\alpha} +f^{\alpha}_{\beta\gamma}\epsilon_{0}^{\beta}W^{\gamma}_{0\; \bar{M}}=0~,
\end{equation}
generating the global higher spin transformation on $\textbf{M}$.
\begin{equation}\label{2044}
	\delta_{\epsilon_{0}} H=\xi_{0}^{\bar{M}}\partial_{\bar{M}}H=\epsilon_{0}^{\beta}H_{\beta},\;\;\;\;\;\;\;\;\delta_{\epsilon_{0}} H_{\alpha}=\xi_{0}^{\bar{M}}\partial_{\bar{M}}H_{\alpha}=\epsilon_{0}^{\beta}\partial_{\beta}H_{\alpha}.
\end{equation}
(\ref{fc35}) is integrable due to (\ref{204}) with the solution characterized by $\epsilon_{0}^{\alpha}$ at one point. With $\epsilon_{0}$ satisfying (\ref{fc35}), (\ref{204}) is invariant under (\ref{2044}). $	[\epsilon_{0},\epsilon'_{0}]^{\alpha} = f^{\alpha}_{\beta\gamma}\epsilon_{0}^{\gamma}\epsilon'^{\beta}_{0}$. The structure constant is undeformed.

The next step is to impose the suitable rheonomy condition and derive the unfolded equation so that the solution on $\textbf{M}$ is determined by the (on-shell) fields in lower dimensions. In the following, we will consider two kinds of the rheonomy conditions which will make the final dynamics reduce to $4d$ and $3d$ respectively. The former gives a system equivalent to the linearized Vasiliev theory expanded on the background $W^{\alpha}_{0}$, which also has an abelian local gauge symmetry invisible if we only focus on the equation for curvature. The latter comes from the $3d$ free massless scalar field theory at $\partial AdS_{4}$. Since the $3d$ scalar forms the representation of the HS symmetry, it is possible to extend the scalar from $3d$ to (the boundary region of) $\textbf{M}$ with the global HS transformation realized as the isometry transformation.

\subsubsection{The 4d global HS invariant system}

Recall that in section 2, (\ref{D00d11}) and (\ref{D00d}) are obtained. With $O(Z)$ replaced by $H(Z)$, from\footnote{(\ref{1q2}) is the generic expansion, among which, some terms may vanish as is explained in Appendix B.}
\begin{eqnarray}\label{1q2}
\nonumber&& \partial_{0\cdots0 a_{1}\cdots a_{s},b_{1}\cdots b_{s+k}}H\\\nonumber&=& 
\sum^{t=1,2,\cdots,2s+k-2r}_{r=0,2,\cdots,s}G^{c_{1}\cdots c_{2r+t}}_{0\cdots0 a_{1}\cdots a_{s},b_{1}\cdots b_{s+k}} 
\partial_{0,c_{2r+t}}\partial_{0,c_{2r+t-1}}\cdots	\partial_{0 c_{1}\cdots c_{r},c_{r+1}\cdots c_{2r+1}}H
,\\
\end{eqnarray}
\begin{eqnarray}\label{1q2a}
&&\nonumber	\partial_{0\cdots0 a^{1}_{1}\cdots a^{1}_{s_{1}},b^{1}_{1}\cdots b^{1}_{s_{1}+k_{1}}} \cdots		\partial_{0\cdots0 a^{p}_{1}\cdots a^{p}_{s_{p}},b^{p}_{1}\cdots b^{p}_{s_{p}+k_{p}}}H \\ &\sim &   \sum \Lambda (a_{1}\cdots a_{s},b_{1}\cdots b_{s+k})\partial_{0,b_{s+k}} \partial_{0,b_{s+k-1}}\cdots \partial_{0 a_{1}\cdots a_{s},b_{1}\cdots b_{s+1}}H
\end{eqnarray}
for the constant $G$ and $\Lambda$, the suitable rheonomy condition can be taken as 
\begin{equation}\label{ll1ml}
	H_{0\cdots0 a_{1}\cdots a_{s},b_{1}\cdots b_{s+k}} =\sum^{t=1,2,\cdots,2s+k-2r}_{r=0,2,\cdots,s} G^{c_{1}\cdots c_{2r+t}}_{0\cdots0 a_{1}\cdots a_{s},b_{1}\cdots b_{s+k}} H_{[0c_{1}\cdots c_{r},c_{r+1}\cdots c_{2r+1}];c_{2r+2}\cdots c_{2r+t}} 
\end{equation}
for $s$ even, $k$ odd, $r$ even; $	H_{0\cdots0 a_{1}\cdots a_{s},b_{1}\cdots b_{s+k}} =0$ for $s$ odd, $k$ even.
\begin{equation}
	H_{[0c_{1}\cdots c_{r},c_{r+1}\cdots c_{2r+1}];c_{2r+2}\cdots c_{2r+t}}=\partial_{c_{2r+t}}\cdots \partial_{c_{2r+2}}H_{[0c_{1}\cdots c_{r},c_{r+1}\cdots c_{2r+1}]}.
\end{equation}
According to the previous decomposition $\alpha =(A,Q)$, $\partial_{Q}H=H_{Q}=0$, so
\begin{equation}\label{vgyh112}
  \partial_{Q}H_{A}=-f^{B}_{QA}H_{B}.
\end{equation}
The evolution along the $Q$ direction is a gauge transformation. The rest Bianchi identity is 
\begin{equation}
	 \partial_{A}H_{B}=\partial_{B}H_{A},  
\end{equation}
which is of course satisfied since (\ref{ll1ml}) is obtained from the scalar operator $O(Z)$ on $\textbf{M}$.

Based on (\ref{1q2}) and (\ref{1q2a}), one may get the unfolded equation 
\begin{eqnarray}\label{sedg}
  \nonumber &&	\partial_{\alpha}H_{[0a_{1} \cdots a_{s},b_{1}\cdots b_{s+1}];c_{1}\cdots c_{n}}=h_{[0a_{1} \cdots a_{s},b_{1}\cdots b_{s+1}];c_{1}\cdots c_{n}\alpha}
   \\\nonumber &&\partial_{\alpha}H_{c_{1}\cdots c_{n}}=h_{c_{1}\cdots c_{n}\alpha}\\\nonumber &\Leftrightarrow & dH_{[0a_{1} \cdots a_{s},b_{1}\cdots b_{s+1}];c_{1}\cdots c_{n}}=h_{[0a_{1} \cdots a_{s},b_{1}\cdots b_{s+1}];c_{1}\cdots c_{n}\alpha}W^{\alpha}_{0} \\ && dH_{c_{1}\cdots c_{n}}=h_{c_{1}\cdots c_{n}\alpha}W^{\alpha}_{0},
\end{eqnarray} 
where $h_{[0a_{1} \cdots a_{s},b_{1}\cdots b_{s+1}];c_{1}\cdots c_{n}\alpha}$ and $h_{c_{1}\cdots c_{n}\alpha}$ are functions of $\{H_{c_{1}\cdots c_{n}},H_{[0a_{1} \cdots a_{s},b_{1}\cdots b_{s+1}];c_{1}\cdots c_{n}}|n=0,1,\cdots;s=2,4,\cdots\}$. So the value of $(H_{[0a_{1} \cdots a_{s},b_{1}\cdots b_{s+1}];c_{1}\cdots c_{n}},H_{c_{1}\cdots c_{n}} )$ at one point determines its value on $\textbf{M}$. Alternatively, $(H,H_{[0a_{1} a_{2},b_{1}b_{2} b_{3}]},H_{[0a_{1} \cdots a_{4},b_{1}\cdots b_{5}]},\cdots)$ on $AdS_{4}$ determines its value on $\textbf{M}$.

The complete $H_{\alpha}$ is exhausted by $H_{Q}=0$ and (\ref{ll1ml}) for $H_{A}$. One may also add $H_{0\cdots 0a_{1}\cdots a_{s},b_{1}\cdots b_{s+k}}$ with $s$ even, $k$ even, which, together with $H_{A}$, forms the twisted-adjoint representation of the higher spin algebra. According to (\ref{b11}), for $s$ even, $H_{[0 a_{1}\cdots a_{s},b_{1}\cdots b_{s+1}]}$ and $H_{[a_{1}\cdots a_{s},b_{1} \cdots b_{s}]}$ are related via 
\begin{equation}
H_{[0 a_{1}\cdots a_{s},b_{1}\cdots b_{s+1}]}=	 \sum_{\{b_{1}\cdots b_{s+1}\}} \partial_{b_{s+1}} H_{[a_{1}\cdots a_{s},b_{1} \cdots b_{s}]}+ \cdots
\end{equation}
So $(H,H_{[0a_{1} a_{2},b_{1}b_{2} b_{3}]},\cdots)$ in $AdS_{4}$ is also equivalent to the field $(H,H_{[a_{1} a_{2},b_{1}b_{2}]}, \cdots)$, which is an irreducible representation of $G[ho(1|2:[3,2])]$.

The relation (\ref{ll1ml}) is obtained from the operator $O(Z)$ on $M$. We may get the similar relation from the linearized Vasiliev theory, where $H_{[a_{1} \cdots a_{s},b_{1}\cdots b_{s}]}\sim R^{s}_{a_{1} \cdots a_{s},b_{1}\cdots b_{s}}$ gets the interpretation as the linearized curvature. Consider 
\begin{eqnarray}
\label{r1}&& 	d W_{0}^{\alpha} =\frac{1}{2}f^{\alpha}_{\beta\gamma}W_{0}^{\beta}\wedge W_{0}^{\gamma} \\ \label{r2}&&
d \Phi^{\tilde{\alpha}}= k^{\tilde{\alpha}}_{\beta\tilde{\gamma}}\Phi^{\tilde{\gamma}} W_{0}^{\beta} \Leftrightarrow\partial_{\beta} \Phi^{\tilde{\alpha}}=k^{\tilde{\alpha}}_{\beta\tilde{\gamma}} \Phi^{\tilde{\gamma}}  
   \\\label{free} & & d\tilde{W}^{\alpha}-f^{\alpha}_{\beta\gamma}W^{\beta}_{0}\wedge \tilde{W}^{\gamma}=\frac{1}{2}R^{\alpha}_{1\beta\gamma}(\Phi^{\tilde{\sigma}})W^{\beta}_{0}\wedge W^{\gamma}_{0}
\end{eqnarray} 
which is the linearized version of the Vasiliev equation (\ref{3w11}) expanded on background $W_{0}^{\alpha}$ with $\tilde{W}^{\alpha}$ the fluctuation on it. $k^{\tilde{\alpha}}_{\beta\tilde{\gamma}} =\rho^{\tilde{\alpha}}_{\alpha}\rho^{\gamma}_{\tilde{\gamma}} f^{\alpha}_{\beta\gamma}$ is a constant. $k^{\tilde{\alpha}}_{\beta\tilde{\gamma}} \Phi^{\tilde{\gamma}}$ is the lowest order term of $F^{\tilde{\alpha}}_{\beta}(\Phi^{\tilde{\sigma}})=k^{\tilde{\alpha}}_{\beta\tilde{\gamma}}(\Phi^{\Phi^{\tilde{\sigma}}}) \Phi^{\tilde{\gamma}}$ in (\ref{3w1}). $R^{\alpha}_{1\beta\gamma}$ is the first order term of the polynomial $\bar{f}^{\alpha}_{\beta\gamma}(\Phi^{\tilde{\sigma}})$ in (\ref{3w1}), i.e. $\bar{f}^{\alpha}_{\beta\gamma}=f^{\alpha}_{\beta\gamma}+R^{\alpha}_{1\beta\gamma}+\mathcal{O}(\Phi^{2})$.

(\ref{r1})-(\ref{free}) are consistent if 
\begin{eqnarray}
\label{sat}&& \frac{\partial R^{\sigma}_{1[\alpha\beta}}{\partial \Phi^{\tilde{\rho}}}k^{\tilde{\rho}}_{\gamma]\tilde{\chi}}\Phi^{\tilde{\chi}}
-f^{\sigma}_{\rho[\alpha}R^{\rho}_{1\beta\gamma]}-R^{\sigma}_{1\rho[\alpha}f^{\rho}_{\beta\gamma]}=0 
   \\\label{sat6} & & k^{\tilde{\alpha}}_{\sigma \tilde{\gamma}}k^{\tilde{\gamma}}_{\rho \tilde{\beta}}-k^{\tilde{\alpha}}_{\rho \tilde{\gamma}}k^{\tilde{\gamma}}_{\sigma \tilde{\beta}}+f^{\chi}_{\rho\sigma}k^{\tilde{\alpha}}_{\chi\tilde{\beta}}=0
\end{eqnarray} 
which is indeed the case due to the vanishing of the the first order part of the left hand side of (\ref{6gy7}).

(\ref{r2})-(\ref{free}) are invariant under the global HS transformation generated by $\xi^{\bar{N}}_{0}=\epsilon^{\alpha}_{0}W^{\bar{N}}_{0\; \alpha}$ preserving the background $W^{\alpha}_{0}$.
\begin{equation}
	d \epsilon_{0}^{\alpha} +f^{\alpha}_{\beta\gamma}\epsilon_{0}^{\beta}W^{\gamma}_{0}=0
\end{equation}
as is in (\ref{fc35}). 
\begin{eqnarray}\label{GLO}
 \nonumber &&	\delta W^{\alpha}_{0}=0,\;\;\;\;\;\;\;\;\;\; \delta \Phi^{\tilde{\alpha}}=\xi^{\bar{N}}_{0}\partial_{\bar{N}}\Phi^{\tilde{\alpha}}= k^{\tilde{\alpha}}_{\beta\tilde{\gamma}} \Phi^{\tilde{\gamma}}\epsilon^{\beta}_{0},
 \\\nonumber&&	\delta \tilde{W}^{\alpha}_{\bar{M}}=\xi^{\bar{N}}_{0}\partial_{\bar{N}}\tilde{W}^{\alpha}_{\bar{M}}+\partial_{\bar{M}}\xi^{\bar{N}}_{0}\tilde{W}^{\alpha}_{\bar{N}},\;\;\;\;\;\;\; \delta R^{\alpha}_{1\rho\sigma}=\xi^{\bar{N}}_{0}\partial_{\bar{N}}R^{\alpha}_{1\rho\sigma}=\frac{\partial R^{\alpha}_{1\rho\sigma}}{\partial \Phi^{\tilde{\alpha}}} k^{\tilde{\alpha}}_{\beta\tilde{\gamma}} \Phi^{\tilde{\gamma}}\epsilon^{\beta}_{0}.\\
\end{eqnarray}
There is also a residue local HS transformation 
\begin{equation}\label{LOC}
	\delta \tilde{W}^{\alpha}=d\epsilon^{\alpha}+ f^{\alpha}_{\beta\gamma}\epsilon^{\beta} W_{0}^{\gamma}=D\epsilon^{\alpha},\;\;\;\;\;\;\;\; \delta W^{\alpha}_{0}=0,\;\;\;\;\;\;\;\; \delta \Phi^{\tilde{\alpha}}=0,\;\;\;\;\;\;\;\; \delta R^{\alpha}_{1\rho\sigma}=0,
\end{equation}
which is invisible if we only focus on the equation for the 0-form. Intuitively, it seems that the global HS transformation for $\tilde{W}^{\alpha}$ and $R^{\alpha}_{1\rho\sigma}$ should be $	\delta \tilde{W}^{\alpha}=f^{\alpha}_{\beta\gamma}\epsilon^{\beta}_{0} \tilde{W}^{\gamma}$ and $\delta R^{\alpha}_{1 \rho\sigma}=f^{\alpha}_{\beta\gamma}\epsilon^{\beta}_{0} R^{\gamma}_{1 \rho\sigma}$, which, however, is not consistent with the transformation law of $\Phi^{\tilde{\alpha}}$. The global HS transformation is a diffeomorphism transformation other than a gauge transformation.

Let us first consider (\ref{r2}). $H=\Phi$, $\partial_{\beta} H=\partial_{\beta} \Phi=k_{\beta\tilde{\gamma}} \Phi^{\tilde{\gamma}}=H_{\beta}$. In particular, 
\begin{equation}\label{4111}
	\partial_{A} H=\partial_{A} \Phi=k_{A\tilde{\gamma}} \Phi^{\tilde{\gamma}}=H_{A}=\Phi_{A},  \;\;\;\;\;\;\;\;\;\;   \partial_{Q} H=\partial_{Q} \Phi=k_{Q\tilde{\gamma}} \Phi^{\tilde{\gamma}}=H_{Q}=0.
\end{equation}
\begin{equation}\label{fgxs1t}
	\partial_{b} \Phi^{[a_{1}\cdots a_{s+t},b_{1}\cdots b_{s}]} =	k^{[a_{1}\cdots a_{s+t},b_{1}\cdots b_{s}]}_{b \;\;\;\;\; [c_{1}\cdots c_{s+t+1},d_{1}\cdots d_{s}]}\Phi^{[c_{1}\cdots c_{s+t+1},d_{1}\cdots d_{s}]}+k^{[a_{1}\cdots a_{s+t},b_{1}\cdots b_{s}]}_{b \;\;\;\;\; [c_{1}\cdots c_{s+t-1},d_{1}\cdots d_{s}]} \Phi^{[c_{1}\cdots c_{s+t-1},d_{1}\cdots d_{s}]}.
\end{equation}
From (\ref{fgxs1t}), we have
\begin{eqnarray}\label{118}
 \nonumber && \partial_{b} \Phi^{[ a_{1}\cdots a_{s},b_{1}\cdots b_{s}]} =	k^{[a_{1}\cdots a_{s},b_{1}\cdots b_{s}]}_{b \;\;\;\;\; [c_{1}\cdots c_{s+1},d_{1}\cdots d_{s}]}\Phi^{[c_{1}\cdots c_{s+1},d_{1}\cdots d_{s}]},\\\nonumber&& 	\partial_{b} \Phi^{[a_{1}\cdots a_{s+1},b_{1}\cdots b_{s}]} =	k^{[a_{1}\cdots a_{s+1},b_{1}\cdots b_{s}]}_{b \;\;\;\;\;[ c_{1}\cdots c_{s+2},d_{1}\cdots d_{s}]}\Phi^{[ c_{1}\cdots c_{s+2},d_{1}\cdots d_{s}]}+k^{[a_{1}\cdots a_{s+1},b_{1}\cdots b_{s}]}_{b \;\;\;\;\; [c_{1}\cdots c_{s},d_{1}\cdots d_{s}]}\Phi^{[c_{1}\cdots c_{s},d_{1}\cdots d_{s}]},\\\nonumber &&  	\partial_{b} \Phi^{[a_{1}\cdots a_{s+2},b_{1}\cdots b_{s}]} =	k^{[a_{1}\cdots a_{s+2},b_{1}\cdots b_{s}]}_{b \;\;\;\;\; [ c_{1}\cdots c_{s+3},d_{1}\cdots d_{s}]}\Phi^{[c_{1}\cdots c_{s+3},d_{1}\cdots d_{s}]}+ k^{[ a_{1}\cdots a_{s+2},b_{1}\cdots b_{s}]}_{b \;\;\;\;\; [c_{1}\cdots c_{s+1},d_{1}\cdots d_{s}]} \Phi^{[c_{1}\cdots c_{s+1},d_{1}\cdots d_{s}]},\\&&\cdots\cdots
\end{eqnarray}
so
\begin{eqnarray}\label{fgxs11a}
 \nonumber &&\partial^{[a(s+1),b(s)]} \Phi = \Phi^{[a(s+1),b(s)]} \sim \partial^{b} \Phi^{[a(s),b(s)]}  ,\\\nonumber&& \partial^{[a(s+2),b(s)]} \Phi =	\Phi^{[a(s+2),b(s])} \sim \partial^{b}\partial^{b} \Phi^{[a(s),b(s)]} +\Phi^{[a(s),b(s)]}  ,\\\nonumber&&	\partial^{[a(s+3),b(s)]} \Phi =\Phi^{[a(s+3),b(s)]} \sim \partial^{b}\partial^{b}\partial^{b} \Phi^{[a(s),b(s)]} +\partial^{b}\Phi^{[a(s),b(s)]} ,\\&&\cdots\cdots
\end{eqnarray}
Compared with the previous discussion on $H_{\alpha}$, $H_{[a_{1}\cdots a_{s},b_{1}\cdots b_{s}]}$ can be identified with $\Phi^{[a(s),b(s)]}$. (\ref{fgxs11a}) is also obtained in \cite{gbfb9vn} by considering the 0-th level unfolded equation of Vasiliev theory, which is just (\ref{r2}).

In the interacting theory, $\partial_{\beta} \Phi^{\tilde{\alpha}}=\hat{k}^{\tilde{\alpha}}_{\beta\tilde{\gamma}}(\Phi^{\tilde{\sigma}}) \Phi^{\tilde{\gamma}}$, 
\begin{eqnarray}\label{eds2a}
 \nonumber && 	\partial_{b_{1}} \Phi^{[a(s),b(s)]}=\hat{k}^{[a(s),b(s)]}_{b_{1} \tilde{\gamma}} \Phi^{\tilde{\gamma}} ,\\\nonumber&& 		\partial_{b_{2}}	\partial_{b_{1}} \Phi^{[a(s),b(s)]}=\frac{\partial \hat{k}^{[a(s),b(s)]}_{b_{1} \tilde{\gamma}}}{\partial \Phi^{\tilde{\sigma}}}	\hat{k}^{\tilde{\sigma}}_{b_{2} \tilde{\rho}} \Phi^{\tilde{\rho}} \Phi^{\tilde{\gamma}}+ \hat{k}^{[a(s),b(s)]}_{b_{1} \tilde{\gamma}} \hat{k}^{\tilde{\gamma}}_{b_{2} \tilde{\rho}} \Phi^{\tilde{\rho}},\\&&\cdots\cdots
\end{eqnarray}
From (\ref{eds2a}), $\{\Phi^{\tilde{\alpha}}\sim \Phi^{[a(s),b(s+k)]}\}$ can be expressed in terms of $\{\partial_{b_{k}}\cdots	\partial_{b_{1}} \Phi^{[a(s),b(s)]}\}$, in a complicated way.

The interpretation of $\Phi^{[a(s),b(s)]}$ as the linearized curvature comes from (\ref{free}). For the background geometry $W_{0}^{\alpha}$ in $\textbf{M}$, one can always choose a particular gauge so that in $AdS_{4}$, 
\begin{equation}\label{aazz}
	\{W^{\alpha}_{0\;\mu}\}=\{W^{a}_{0\;\mu},W^{(ab)}_{0\;\mu},0,0,\cdots \},
\end{equation}
where $W^{a}_{0\;\mu}$ and $W^{(ab)}_{0\;\mu}$ are the vielbein and the connection charactering $AdS_{4}$ geometry. (\ref{free}) becomes (\ref{387}) with $R^{[a(s-1),b(s-1)]}_{ab} \sim \Phi^{[a(s),b(s)]}$ the linearized Weyl tensor in free higher spin theory. (\ref{r1})-(\ref{free}) indicates that not only the interacting HS theory (Vasiliev theory), the free HS theory (Fronsdal theory) can also be consistently extended to $\textbf{M}$ with the symmetry reducing to the global HS symmetry and an abelian local HS symmetry.

Fronsdal equation for metric-like fields is invariant under the local HS transformation. A natural question is whether there are any manifestations of the global HS symmetry. Note that the ``central on-mass-shell theorem'' is the necessary condition for the linearized Vasiliev equation to reduce to the Fronsdal equation. In Vasiliev theory, 
\begin{equation}
	R^{[a(s-1),b(t)]}_{1ab}(\Phi^{\tilde{\sigma}})=0 \;\;\;\;\;\;\;\;\; \textnormal{for} \;\;\;\;\;\;\;\;\; t \neq s-1
\end{equation}
is valid as a function equation independent of the position in $\textbf{M}$. So the ``central on-mass-shell theorem'' is preserved under the global HS transformation, which is just a diffeomorphism transformation on $\textbf{M}$. Under the global HS transformation, we move from one $AdS$ fiber to another, with the Fronsdal equation satisfied as well. However, the transformation is on-shell and nontrivial, since we must first solve $\tilde{W}^{\alpha}$ all over $\textbf{M}$ and then perform (\ref{GLO}). More explicitly, the transformation law of $\tilde{W}^{\alpha}_{\mu}$ in $AdS_{4}$ is  
\begin{equation}
	\delta \tilde{W}^{\alpha}_{\mu}=\xi^{\bar{N}}_{0}\partial_{\bar{N}}\tilde{W}^{\alpha}_{\mu}+\partial_{\mu}\xi^{\bar{N}}_{0}\tilde{W}^{\alpha}_{\bar{N}}.
\end{equation}
Suppose $\bar{N}=\{\tilde{N},\mu\}$, for simplicity, we may let $\tilde{W}^{\alpha}_{\tilde{N}}=0$ in $AdS_{4}$, then from (\ref{free}),  
\begin{eqnarray}\label{RIG}
 \nonumber  		\delta \tilde{W}^{[a(s-1),b(0)]}_{\mu}&=&\xi^{\tilde{N}}_{0}W^{\beta}_{0\; \tilde{N}}(f^{[a(s-1),b(0)]}_{\beta\gamma}\tilde{W}^{\gamma}_{\mu}-R^{[a(s-1),b(0)]}_{1a \beta}W^{a}_{0\; \mu})\\&+& 		\xi^{\nu}_{0}\partial_{\nu}\tilde{W}^{[a(s-1),b(0)]}_{\mu}+\partial_{\mu}\xi^{\nu}_{0}\tilde{W}^{[a(s-1),b(0)]}_{\nu}.
\end{eqnarray}
$R^{\alpha}_{1 a\beta}(\Phi^{\tilde{\sigma}})=r^{\alpha}_{a\beta}|_{\tilde{\sigma}}\Phi^{\tilde{\sigma}}$ with $r^{\alpha}_{\beta\gamma}|_{\tilde{\sigma}}$ the constant. $\Phi^{\tilde{\sigma}}$ can be expressed in terms of the $4d$ derivatives of $\Phi^{[a(s),b(s)]}$, which, in turn, is determined by $\tilde{W}^{\alpha}_{\mu}$ and thus $\tilde{W}^{[a(s-1),b(0)]}_{\mu}$. The right hand side of (\ref{RIG}) is a complicated $4d$ linear differential operator on $\tilde{W}^{[a(s-1),b(0)]}_{\mu}$. However, $\delta \tilde{W}^{[a(s-1),b(0)]}_{\tilde{N}} \neq 0$, the simplification condition $\tilde{W}^{\alpha}_{\tilde{N}}=0$ is not preserved. In contrast to the gauge field, the global HS transformation law of the Weyl tensor in Fronsdal theory is straightforward.  
\begin{equation}
	\delta \Phi^{[a(s),b(s)]}=k^{[a(s),b(s)]}_{\beta \tilde{\gamma}}\Phi^{\tilde{\gamma}}\epsilon^{\beta}_{0},
\end{equation}
where $\Phi^{\tilde{\gamma}}$ can be written in terms of the $4d$ derivatives of $\Phi^{[a(s),b(s)]}$ via the relation $D_{\mu}\Phi^{\tilde{\alpha}}=k^{\tilde{\alpha}}_{a \tilde{\gamma}}\Phi^{\tilde{\gamma}}W^{a}_{0\;\mu}$.

It is well-known that the linearized Vasiliev theory is global HS invariant. By extending the space from $AdS_{4}$ to $\textbf{M}$, the linearized Weyl module of the free higher spin theory can be compactly interpreted as $\partial_{\alpha}H$, the outer derivatives of a single scalar field $H$ on $\textbf{M}$.

\subsubsection{The 3d global HS invariant system}

The above $4d$ global HS invariant theory also has a local gauge symmetry. The genuine global HS invariant system without the local gauge symmetry is the $3d$ massless free scalar field theory living at $\partial AdS_{4}$. In $3d$ free CFT, let $\phi$ be the operator for the dimension $1/2$ massless scalar and consider $\phi(X)=g(X) \phi(0')g(X)^{-1}$, $\forall\; g(X) \in G[ho(1|2:[3,2])]$. In contrast to $O(0)$ in the bulk, $\phi(0')$ is at the origin of $\partial AdS_{4}$, so for the finite $X$, $\phi(X)$ is still at the near boundary region of $\textbf{M}$ with $X$ the coordinate.

Scalar field at the near boundary region of $\textbf{M}$ also forms the representation of $G[ho(1|2:[3,2])]$. $-i \partial_{\alpha}\phi(X)=[Q_{\alpha}(X),\phi(X)]$. For $Q \in so(3,2)$,  
\begin{eqnarray}
  \nonumber &&	[K_{m }(0'),\phi(0')]=0,\;\;\;\;\;[P_{m }(0'),\phi(0')]=-i \partial_{m}\phi(0'), \\ && 	[Q_{m,n}(0'), \phi(0')]=0, \;\;\;\;\;[Q_{0,4}(0'),\phi(0')]=-\frac{i}{2}\phi(0').
\end{eqnarray} 
Generically, in $3d$ free CFT of the scalar $\phi$, we have the relation  
\begin{equation}
 \partial_{\alpha}\phi(0')=i[Q_{\alpha}(0'),\phi(0')] = \sum_{k}(-i)^{k}\rho_{\alpha}^{i_{1}\cdots i_{k}}[P_{i_{1}}(0'),\cdots[P_{i_{k}}(0'),\phi(0')]\cdots], 
\end{equation}
where $i_{k}=1,2,3$, $\rho$ is the constant, because $ho(1|2:[3,2])$ can be realized as the quotient of the enveloping algebra of $so(3,2)$ \cite{vas1} (for HS algebra of any classical Lie algebras and in particular, $\mathfrak{sp}_{2N}$, $\mathfrak{so}_{N}$ and $\mathfrak{sl}_{N}$, see \cite{vas12}). As a result, the relation 
\begin{equation}\label{iml}
 \partial_{\alpha}\phi(X) = \sum_{k}(-i)^{k}\rho_{\alpha}^{i_{1}\cdots i_{k}}[P_{i_{1}}(X),\cdots[P_{i_{k}}(X),\phi(X)]\cdots]=\sum_{k}\rho_{\alpha}^{i_{1}\cdots i_{k}}\partial_{i_{k}}\cdots \partial_{i_{1}}\phi(X)
\end{equation}
is valid everywhere at the boundary region of $\textbf{M}$ for the constant $\rho$. $3d$ equations of motion for $\phi$ are also implicitly imposed by (\ref{iml}). The derivatives of $\phi$ in outer space can be expressed in terms of the derivatives of $\phi$ in inner space ($\partial AdS_{4}$). This is not possible in (\ref{1q2}), because the scalar field in $AdS_{4}$ cannot form the representation of the HS symmetry. One must introduce the higher spin fields, which, in (\ref{1q2}), is reflected by $\partial_{0 c_{1}\cdots c_{r},c_{r+1}\cdots c_{2r+1}}$.

Return to (\ref{204})-(\ref{2044}) and restrict to the near boundary region with $H$ replaced by $\phi$. From the rheonomy condition
\begin{equation}\label{3114}
	\phi_{\alpha}=\sum_{k}\rho_{\alpha}^{i_{1}\cdots i_{k}}\partial_{i_{k}}\cdots \partial_{i_{1}} \phi=\sum_{k}\rho_{\alpha}^{i_{1}\cdots i_{k}}\phi_{i_{1}\cdots i_{k}},
\end{equation}
one may get the unfolded equation 
\begin{equation}\label{GL}
	\partial_{\alpha}\phi_{i_{1}\cdots i_{n}}=\phi_{i_{1}\cdots i_{k}\alpha}\Leftrightarrow d\phi_{i_{1}\cdots i_{n}}=\phi_{i_{1}\cdots i_{k}\alpha}W^{\alpha}_{0},
\end{equation}
where $\phi_{i_{1}\cdots i_{k}\alpha}$ is the linear combination of $\{\phi,\phi_{i_{1}},\phi_{i_{1}i_{2}},\cdots\}$ with the constant coefficients.

The Bianchi identity 
\begin{equation}
	\frac{\partial \phi_{i_{1}\cdots i_{k}\gamma}}{\partial \phi_{j_{1}\cdots j_{n}}}\phi_{j_{1}\cdots j_{n}\beta}-\frac{\partial \phi_{i_{1}\cdots i_{k}\beta}}{\partial \phi_{j_{1}\cdots j_{n}}}\phi_{j_{1}\cdots j_{n}\gamma}+f^{\alpha}_{\beta\gamma}\phi_{i_{1}\cdots i_{k}\alpha}=0
\end{equation}
is satisfied. From the on-shell $\{\phi,\phi_{i_{1}},\phi_{i_{1}i_{2}},\cdots\}$ at one point, or equivalent, the on-shell $\phi$ in $\partial AdS_{4}$, $\phi$ in the near boundary region of $\textbf{M}$ can be determined. (\ref{GL}) is invariant under the global HS transformation
\begin{equation}
	\delta_{\epsilon_{0}} \phi_{i_{1}\cdots i_{n}}=\xi^{\bar{N}}_{0}\partial_{\bar{N}}\phi_{i_{1}\cdots i_{n}}= \epsilon^{\alpha}_{0}\phi_{i_{1}\cdots i_{n}\alpha}.
\end{equation}

In conclusion, to construct a theory with the global HS symmetry, we may try to find an unfolded equation like (\ref{sedg}) and (\ref{GL}) for a 0-form multiplet on $\textbf{M}$ with the background geometry $W^{\alpha}_{0}$. The equation should be integrable with the only dependence on $\textbf{M}$ comes from the 0-form and the 1-form $W^{\alpha}_{0}$. Therefore, it is of course diffeomorphism invariant. The global higher spin transformation is a special diffeomorphism transformation preserving $W^{\alpha}_{0}$.

\section{Discussion}

In supergravity, the rheonomy condition is simply $R^{A}_{BC} = r^{A}_{BC}(R^{cd}_{ab},R^{\alpha}_{ab},H,H_{\alpha})$. Nevertheless, the most generic rheonomy condition in group manifold approach takes the form of (\ref{1azd5}) and (\ref{hy}) with all orders of derivatives included. If we make a similar truncation $R^{\alpha}_{\beta\gamma}=r^{\alpha}_{\beta\gamma}(R^{[a(s-1),b(s-1)]}_{ab},H)$ in higher spin theory, then with $r^{\alpha}_{\beta\gamma}$ plugged into the Bianchi identity, we will get the $4d$ equations of motion, which, when expressed in terms of $(W^{[a(s-1),b(0)]}_{\mu},H)$, do not contain derivatives higher than two. However, it is quite likely that such equations may only have the trivial solution $R^{[a(s-1),b(s-1)]}_{ab}=H=0$ no matter how the coefficients in function $r^{\alpha}_{\beta\gamma}$ are adjusted. To allow for the nontrivial on-shell degrees of freedom, higher derivatives must be included so that $R^{\alpha}_{\beta\gamma}$ at one point is effectively determined by $(W^{[a(s-1),b(0)]}_{\mu},H)$ on the whole $AdS_{4}$. The $4d$ equations of motion for $(W^{[a(s-1),b(0)]}_{\mu},H)$ will also contain an infinite number of the higher derivative terms which make the theory nonlocal.

To write the unfolded equations (\ref{nanal}) and (\ref{5g})-(\ref{7g}), the infinite $0$-form multiplets are necessarily involved in both supergravity and higher spin theory, since the solutions on the whole $\textbf{M}$, including $M_{4}$/$AdS_{4}$, are characterized by the on-shell $0$-form multiplets at one point. For higher spin theory, the on-shell $(R^{[a(s-1),b(s-1)]}_{ab;c_{1}\cdots c_{n}},H_{c_{1}\cdots c_{n}})$ is equivalent to $\{\Phi^{[a(s+n),b(s)]}\}$, so the solution on $\textbf{M}$ is also characterized by the arbitrary $\{\Phi^{[a(s+n),b(s)]}\}$ at that point. Merely based on group manifold approach without the knowledge of Vasiliev theory, we will finally arrive at (\ref{1g})-(\ref{9g}) and then face the problem of finding the proper rheonomy condition that could solve the Bianchi identity, allow for the maximum on-shell degrees of freedom and have the correct free theory limit. It is Vasiliev theory that gives the solution meeting all these requirements. A question is whether there are other solutions or not. In Appendix C, we give a rheonomy condition (for the bosonic higher spin theory) satisfying the Bianchi identity with the on-shell degrees of freedom $\{\Phi^{\alpha}\}$. However, the correct free theory limit is not recovered and the local Lorentz transformation is deformed.

In superspace with the fixed background geometry, the local super Poincare symmetry reduces to the global super Poincare symmetry. With the chiral constraint imposed, the component expansion of the scalar superfield in superspace gives the spin $0$ and $1/2$ fields $(H,H^{\alpha})$ in $M_{4}$. For higher spin theory, one can fix the background geometry of $\textbf{M}$ and then study the scalar field $H$ in $\textbf{M}$ with the global higher spin symmetry. The component expansion of $H$ gives the spin $0,2,4,\cdots$ fields $(H,H_{[a_{1}a_{2},b_{1}b_{2}]},H_{[a_{1}a_{2}a_{3}a_{4},b_{1}b_{2}b_{3}b_{4}]},\cdots)$ in $AdS_{4}$, which, however, are not the gauge fields but the linearized Weyl tensors of the free HS theory, since the massless gauge fields are not the Lorentz tensor. Restricted to the near boundary region of $\textbf{M}$, it is also possible to impose the rheonomy constraint so that the component expansion of $H$ only gives the spin $0$ field $H$ in $\partial AdS_{4}$. This is because although the $4d$ spin $0,2,4,\cdots$ fields all together form the representation of the HS symmetry, the $3d$ spin $0$ field alone forms the HS representation.

\bigskip
\bigskip

{\bf Acknowledgments:}

This research was supported in part  by the Natural Science
Foundation of China under grant numbers 10821504, 11075194, 11135003, 11275246, and 11475238,
and by the National Basic Research Program of China (973 Program) under grant number 2010CB833000.

\appendix

\section{$G[ho(1|2:[3,2])]/E$ from the CFT operators}

The minimal bosonic higher spin theory in $AdS_{4}$ is dual to the $3d$ $O(N)$ vector model \cite{hc,hf}. The conserved charges in both theories form the algebra isomorphic to $ho(1|2:[3,2])$. The duality requires that the states and the operators in CFT and the bulk theory can be identified, so in the following, we will use the operators in CFT to represent their counterparts in $4d$ HS theory. Suppose $\{Q_{\alpha}\sim Q_{A_{1}\cdots A_{s-1},B_{1}\cdots B_{s-1}}\}$ are charge operators in CFT corresponding to $\{t_{\alpha}\sim t_{A_{1}\cdots A_{s-1},B_{1}\cdots B_{s-1}}\}$ in (\ref{labd}). $A_{k},B_{k}=0,1,2,3,4$. The explicit form of $Q_{\alpha}$ can be found in \cite{cfjy1}. The CFT realization of the bulk local field operators is usually constructed as \cite{h1, h2, h3} 
\begin{equation}
	\Phi (x) \leftrightarrow \int dX \; K(X|x)\; \mathcal{O}(X)
\end{equation}
in large $N$ limit, where $\Phi (x)$ is the bulk field in $AdS$, $\mathcal{O}(X)$ is the boundary operator in CFT, $K(X|x)$ is the boundary-bulk propagator. $\Phi (x)$ like this of course satisfies the free field equation in $AdS$, which is acceptable when $N\rightarrow \infty$. In this section and the next one, we will construct the spin $0$ field operator and the spin $s$ linearized curvature tensor operators in $AdS$ for $s=2,4,\cdots$, using the CFT operators $O_{i_{1}\cdots i_{s}}$ in \cite{cfjy1}. $i_{k}=1,2,3$. $O_{i_{1}\cdots i_{s}}(x)$ only contains the positive frequency part, but it is enough for the present use.

For $AdS_{4}$ parameterized by $x^{2}_{0}-x^{2}_{1}-x^{2}_{2}-x^{2}_{3}+ x^{2}_{4} = 1$, let $Q_{A,B}$ be the generators of $SO(3,2)$, then for an operator $O(x)$ with $s=0$,
\begin{equation}\label{a1}
		[Q_{A,B}, O(x)]= i(x_{A}\partial_{B}-x_{B}\partial_{A})O(x). 
\end{equation}
Without losing of the generality, let us consider a point $0$ in the bulk of $AdS_{4}$ with the coordinate $x^{0}=1$, $x^{1}=\cdots=x^{4}=0$.
\begin{equation}\label{90ok}
	[Q_{m,n}, O(0)]=[Q_{m, 4}, O(0)]=[K_{m }-iP_{m }, O(0)]=0,  
\end{equation}
where $m,n =1,2,3$. $\{Q_{0,4}, Q_{0,m}\} \subset K(0)$ generates the tangent space along $AdS_{4}$. From (\ref{90ok}), according to the operators constructed in \cite{cfjy1}, $O(0)$ is solved as 
\begin{equation}\label{238}
O(0) = \sum \frac{k!}{(2k+1)!!}	 a^{+}_{i_{1}\cdots i_{k}}a^{+}_{i_{1}\cdots i_{k}}. 
\end{equation}
In \cite{cfjy1}, the generic elements of $ho(1|2:[3,2])$ can be written as 
\begin{equation}\label{eq4}
	Q_{m_{1}\cdots m_{p},n_{1}\cdots n_{q}} =i \sum g(l) \; a^{+}_{m_{1}\cdots m_{p}i_{1}\cdots i_{l}}a_{n_{1}\cdots n_{q}i_{1}\cdots i_{l}} 
\end{equation}
with $m_{k},n_{k},i_{k}=1,2,3$, so
\begin{equation}
	[	Q_{m_{1}\cdots m_{p},n_{1}\cdots n_{q}}, O(0)] \sim i \sum g(l) \; a^{+}_{m_{1}\cdots m_{p}i_{1}\cdots i_{l}}a^{+}_{n_{1}\cdots n_{q}i_{1}\cdots i_{l}}. 
\end{equation}
One can choose the basis $\{Q\}$ of $ho(1|2:[3,2])$ with the definite conformal dimension. 
\begin{equation}
	[D, Q]= -i \Delta Q, \;\;\;\;\;\;\;\;\;\;[D, Q^{+}]= i \Delta Q^{+}.
\end{equation}
Let $H_{Q} = Q+Q^{+}$, $\bar{H}_{Q} =i( Q-Q^{+})$, there will be   
\begin{equation}
	[H_{Q}, O(0)]=0,\;\;\;\;\;\;\;\;\;\;[\bar{H}_{Q}, O(0)]=2i[Q, O(0)].
\end{equation}
$a[E(0)] = \{H_{Q}\}$, $K(0)=\{\bar{H}_{Q}\} $. Moreover, 
\begin{equation}\label{42}
	[\{\bar{H}_{Q}\}, \{\bar{H}_{Q}\}]\subset \{H_{Q}\}, \;\;\;\;\;\;\; 	[\{H_{Q}\}, \{H_{Q}\} ]\subset \{H_{Q}\}, \;\;\;\;\;\;\; 	[\{H_{Q}\}, \{\bar{H}_{Q}\} ]\subset \{\bar{H}_{Q}\}.
\end{equation}
$M$ is a symmetric space.

For $ho(1|2:[3,2])$, there is an involution $\sigma$ 
\begin{equation}
	\sigma (Q) = Q^{+}
\end{equation}
with $\sigma^{2}=1$. $\sigma$ has the eigenvalues $1$ and $-1$ with $\{H_{Q}\}$ and $\{\bar{H}_{Q}\}$ defined above the corresponding eigenspaces. 
\begin{equation}
	ho(1|2:[3,2]) = \{H_{Q}\} \oplus \{\bar{H}_{Q}\}. 
\end{equation}
Under the Wick rotation, $x^{0} \rightarrow ix^{0}$, the action of $\sigma$ is then $\sigma :\; ix^{0}\rightarrow -ix^{0}$, so 
\begin{eqnarray}
\nonumber && a[E(0)] =\{H_{Q}\} = 	\{t_{0 \cdots 0a_{1},b_{1}\cdots b_{s-1}},t_{0\cdots 0a_{1}a_{2} a_{3},b_{1}\cdots b_{s-1}},\cdots, t_{a_{1}\cdots a_{s-1},b_{1}\cdots b_{s-1}}\},
 \\ && K(0)=\{\bar{H}_{Q}\} = 	\{t_{0\cdots 0,b_{1}\cdots b_{s-1}},t_{0\cdots 0a_{1} a_{2},b_{1}\cdots b_{s-1}},\cdots, t_{0a_{1}\cdots a_{s-2},b_{1}\cdots b_{s-1}}\}.
\end{eqnarray}
The decomposition is consistent with (\ref{ae}) and (\ref{aee}).

\section{CFT realization of the spin $s$ linearized Riemann tensor operator in $AdS_{4}$}

In radial quantization of the $3d$ $O(N)$ vector model, for each $s=0,2,\cdots$, there is an unique primary operator $O_{i_{1}\cdots i_{s}}(0')$ with spin $s$.
\begin{equation}\label{ses}
\frac{1}{2}[Q^{A,B},[Q_{A,B}, O_{i_{1}\cdots i_{s}}(0')]]	\equiv[C_{2}, O_{i_{1}\cdots i_{s}}(0')]= 2(s^{2}-1) O_{i_{1}\cdots i_{s}}(0'), 
\end{equation}
where $C_{2}$ is the Casimir operator. $i_{k}=1,2,3$. Here $0'$ represents the origin in the boundary CFT and should be distinguished from the $0$ in Appendix A. $\{\partial_{\mu_{1}}\cdots \partial_{\mu_{n}}O_{i_{1}\cdots i_{s}}(0')|s=0,2,\cdots;n=0,1,\cdots\}$ forms the complete basis of the 1-particle Hilbert space. $\mu_{k}=1,2,3$. $\{O(0'),O_{i_{1}i_{2}}(0'),\cdots\}$ is the higher spin multiplet. The action of the generic $Q_{\alpha}$ on the spin $0$ primary operator $O(0')=a^{+}a^{+}$ can be decomposed as 
\begin{equation}\label{DEC}
		[Q_{A_{1}\cdots A_{s-1},B_{1}\cdots B_{s-1}},O(0')]= \sum^{t=0,1,\cdots}_{r=0,2,\cdots} g^{\mu_{1}\cdots \mu_{t};i_{1}\cdots i_{r}}_{A_{1}\cdots A_{s-1},B_{1}\cdots B_{s-1}} \partial_{\mu_{1}}\cdots \partial_{\mu_{t}}O_{i_{1}\cdots i_{r}}(0').
\end{equation}

Let us construct the $SO(3,1)$ tensor operator with spin $s$ in the sense of (\ref{ses}) in $AdS$ bulk. Such operator does not represent the spin $s$ gauge field which is not a tensor, but rather the field strength of it. The spin $0$ operator $O(0)$ is already given by (\ref{238}). For operators with the higher spin, consider 
\begin{eqnarray}\label{yw}
 A^{+}_{m_{1}m_{2} ,k}&=& a^{+}_{m_{1}m_{2} i_{1}\cdots i_{k} }a^{+}_{i_{1}\cdots i_{k}}+f_{1}a^{+}_{m_{1} i_{1}\cdots i_{k} }a^{+}_{m_{2}  i_{1}\cdots i_{k}}         ,    \nonumber \\ A^{+}_{m_{1}m_{2}m_{3}m_{4} ,k}&=&	a^{+}_{m_{1}m_{2}m_{3}m_{4} i_{1}\cdots i_{k} }a^{+}_{i_{1}\cdots i_{k}}+f_{1}a^{+}_{m_{1}m_{2}m_{3}i_{1}\cdots i_{k}  }a^{+}_{m_{4}i_{1}\cdots i_{k}}+\cdots \nonumber \\&& +f_{4}a^{+}_{m_{2}m_{3}m_{4}i_{1}\cdots i_{k}  }a^{+}_{m_{1}i_{1}\cdots i_{k}} + f_{5} a^{+}_{m_{1}m_{2} i_{1}\cdots i_{k}  }a^{+}_{m_{3}m_{4}i_{1}\cdots i_{k}}
 \nonumber \\&&+ f_{6} a^{+}_{m_{1}m_{3} i_{1}\cdots i_{k}  }a^{+}_{m_{2}m_{4}i_{1}\cdots i_{k}}+f_{7}a^{+}_{m_{1}m_{4} i_{1}\cdots i_{k}  }a^{+}_{m_{2}m_{3}i_{1}\cdots i_{k}},\nonumber \\\cdots &&
\end{eqnarray}
which is the most generic $s$-tensor with the dimension $s+2k+1$. $m_{p},i_{p}=1,2,3$. For each $s$, imposing the condition  
\begin{equation}
	[C_{2}, A^{+}_{m_{1}\cdots m_{s} ,k}]= 2(s^{2}-1) A^{+}_{m_{1}\cdots m_{s}, k}
\end{equation}
can uniquely fix $f_{i}$ in (\ref{yw}). The corresponding operator is denoted as $A^{(s)+}_{m_{1}\cdots m_{s} ,k}$, which is totally symmetric and traceless.

Suppose 
\begin{equation}\label{345}
O^{s}_{m_{1}\cdots m_{s}}(0) = \sum  g(k)    A^{(s)+}_{m_{1}\cdots m_{s} ,k}
\end{equation}
is a spin $s$ tensor operator at $0$, then in analogy with (\ref{90ok}),\footnote{$O^{s}_{m_{1}\cdots m_{s}}(0)$ is a gauge invariant operator. The $SO(3,1)$ transformation of the spin $s$ massless gauge field also has the derivative terms on the right hand side.} 
\begin{equation}\label{sup}
		[Q_{m,n}, O^{s}_{m_{1}\cdots m_{s}}(0)]=\Sigma_{mn} O^{s}_{m_{1}\cdots m_{s}}(0),\;\;\;\;\;\;\;\;\;	[Q_{4,m}, O^{s}_{m_{1}\cdots m_{s}}(0)]=\Sigma_{4m} O^{s}_{m_{1}\cdots m_{s}}(0),
\end{equation}
$m,n=1,2,3$. $\Sigma$ is the spin operator. The first equation in (\ref{sup}) is satisfied for the arbitrary $g(k)$. $O^{s}_{m_{1}\cdots m_{s}}(0)$ forms the representation of $SO(3)$. The complete $SO(3,1)$ representation can be obtained by the successive action of $Q_{4,m}$. The coefficient $g(k)$ in (\ref{345}) is determined by the requirement that at some point, no new operators can be created as is required by the second equation of (\ref{sup}). When $s=0$, the solution of $[Q_{4,m}, O^{0} (0)]=0$ is $O(0)$ in (\ref{238}). When $s=2$, the minimal times for the action of $Q_{4,m}$ is $3$. The corresponding $O^{2}_{m_{1} m_{2}}(0)$ can be written as $O^{2}_{m_{1} m_{2},44}(0)$, while the action of $\{Q_{4,m},Q_{m,n}\}$ gives the complete $SO(3,1)$ representation $O^{2}_{b_{1} b_{2},b_{3} b_{4}}(0)$ with $b_{i}=1,2,3,4$. Generically, for spin $s$ operator $O^{s}_{m_{1}\cdots m_{s}}(0)$, we have $O^{s}_{m_{1}\cdots m_{s}}(0)\equiv O^{s}_{m_{1}\cdots m_{s}, 4\cdots4 }(0)$ with the $SO(3,1)$ completion $O^{s}_{b_{1} \cdots b_{s},b_{s+1} \cdots b_{2s}}(0)$. The maximum number of $4$ in $O^{s}_{b_{1} \cdots b_{s},b_{s+1} \cdots b_{2s}}(0)$ is $s$.

The minimum spin $s$ $SO(3,1)$ tensor operator transforming like (\ref{sup}) is not $O^{s}_{b_{1} \cdots b_{s}}$ but $O^{s}_{b_{1} \cdots b_{s},b_{s+1} \cdots b_{2s}}$. This is expected, since the massless gauge field is not a Lorentz tensor. $O^{s}_{b_{1} \cdots b_{s},b_{s+1} \cdots b_{2s}}$ matches well with the Riemann curvature $R^{s}_{b_{1} \cdots b_{s},b_{s+1} \cdots b_{2s}}$ of the spin $s$ field, or more precisely, the linearized Riemann curvature in $AdS$ background since $O^{s}_{b_{1} \cdots b_{s},b_{s+1} \cdots b_{2s}}$ only creates single particle states.

$\{Q_{0,b}\}$ generates the tangent space at $0$ along $AdS_{4}$. The Successive action of $Q_{0,b}$ gives 
\begin{equation}\label{mal}
	O^{s}_{b_{1} \cdots b_{s},b_{s+1} \cdots b_{2s};b_{2s+1} \cdots b_{2s+k}}(0)= [Q_{0,b_{2s+k}},  \cdots[Q_{0,b_{2s+2}},[Q_{0,b_{2s+1}},O^{s}_{b_{1} \cdots b_{s},b_{s+1} \cdots b_{2s}}(0)]]\cdots]. 
\end{equation}
$	O^{s}_{b_{1} \cdots b_{s},b_{s+1} \cdots b_{2s};b_{2s+1} \cdots b_{2s+k}}(0)$ is the descendant of $O^{s}_{b_{1} \cdots b_{s},b_{s+1} \cdots b_{2s}}(0)$ thus is a spin $s$ operator as well. $\forall \; x \in AdS_{4}$,\footnote{The relation (\ref{ref1w}) is not valid for $O_{i_{1}\cdots i_{s}}$, which is not a tensor.}
\begin{eqnarray}\label{ref1w}
\nonumber &&	 O^{s}_{b_{1} \cdots b_{2s}}(x)=g(x)O^{s}_{b_{1} \cdots b_{2s}}(0)g(x)^{-1},\\ && O^{s}_{b_{1} \cdots b_{s},b_{s+1} \cdots b_{2s};b_{2s+1} \cdots b_{2s+k}}(x)=g(x)O^{s}_{b_{1} \cdots b_{s},b_{s+1} \cdots b_{2s};b_{2s+1} \cdots b_{2s+k}}(0)g(x)^{-1}, 
\end{eqnarray}
$g(x)\in SO(3,2)$. $\{O^{s}_{b_{1} \cdots b_{s},b_{s+1} \cdots b_{2s};b_{2s+1} \cdots b_{2s+k}}(x)|s=0,2,\cdots;k=0,1,\cdots\}$ at $x$ compose the complete basis for the 1-particle Hilbert space of the $4d$ HS theory.

Now consider $[Q_{0\cdots0 a_{1}\cdots a_{s},b_{1}\cdots b_{s+k}},O(0)]$ with $k=1,3,\cdots$, $a_{i},b_{i}=1,2,3,4$, which could be expanded as
\begin{eqnarray}\label{dfr}
\nonumber &&	 	[Q_{0\cdots0 a_{1}\cdots a_{s},b_{1}\cdots b_{s+k}},O(0)]\\ \nonumber&=& \sum^{t=0,1,\cdots,2s+k-2r}_{r=0,2,\cdots,s}  k^{c_{1} \cdots c_{2r+t}}_{a_{1}\cdots a_{s},b_{1}\cdots b_{s+k}}[Q_{0,c_{2r+t}},  \cdots[Q_{0,c_{2r+2}},[Q_{0,c_{2r+1}},O^{r}_{c_{1} \cdots c_{r},c_{r+1} \cdots c_{2r}}(0)]]\cdots].\\
\end{eqnarray}
(\ref{DEC}) could be taken as the boundary limit of (\ref{dfr}), where the linearized curvature tensor has already been written as the derivatives of the metric operator $O_{i_{1}\cdots i_{s}}$. There is no charge operator that could directly create $O^{s}_{a_{1}\cdots a_{s},b_{1} \cdots b_{s}}(0)$ from $O(0)$, the closest one is 
\begin{equation}\label{b11}
	[Q_{0 a_{1}\cdots a_{s},b_{1}\cdots b_{s+1}},O(0)] = \sum_{\{b_{1}\cdots b_{s+1}\}}[Q_{0,b_{s+1}},O^{s}_{a_{1}\cdots a_{s},b_{1} \cdots b_{s}}(0)]+ \cdots.
\end{equation}
$\cdots$ are possible terms with the spin lower than $s$. (\ref{dfr}) and (\ref{b11}) represent the generic possibilities, among which some terms may simply vanish. Since $k$ is odd, instead of the ``primary'' $O^{s}_{a_{1}\cdots a_{s},b_{1} \cdots b_{s}}(0)$, one can also use the less ``primary'' $[Q_{0 a_{1}\cdots a_{s},b_{1}\cdots b_{s+1}},O(0)] $, 
\begin{eqnarray}\label{ccoo}
\nonumber && [Q_{0\cdots0 a_{1}\cdots a_{s},b_{1}\cdots b_{s+k}},O(0)]\\\nonumber
 &=& \sum^{t=1,\cdots,2s+k-2r}_{r=0,2,\cdots,s} f^{c_{1}\cdots c_{2r+t}}_{0\cdots0 a_{1}\cdots a_{s},b_{1}\cdots b_{s+k}} [Q_{0,c_{2r+t}},\cdots [Q_{0,c_{2r+2}},[Q_{0 c_{1}\cdots c_{r},c_{r+1}\cdots c_{2r+1}},O(0)]]\cdots] \\\nonumber
 &=& \sum^{t=1,\cdots,2s+k-2r}_{r=0,2,\cdots,s} g^{c_{1}\cdots c_{2r+t}}_{0\cdots0 a_{1}\cdots a_{s},b_{1}\cdots b_{s+k}} [Q_{0 c_{1}\cdots c_{r},c_{r+1}\cdots c_{2r+1}},\cdots [Q_{0,c_{2r+t-1}},[Q_{0,c_{2r+t}},O(0)]]\cdots].  \\
\end{eqnarray}
Especially, 
\begin{equation}\label{coco}
	 [Q_{0\cdots0,b_{1}\cdots b_{k}},O(0)]\sim \sum_{\{b_{1}\cdots b_{k}\}} [Q_{0,b_{k}} ,\cdots[Q_{0,b_{2}} ,[Q_{0,b_{1}} ,O(0)]]\cdots],
\end{equation}
$[Q_{0\cdots0,b_{1}\cdots b_{k}},O(0)]$ is the descendant of $O(0)$.

\section{A rheonomy condition satisfying the Bianchi identity without giving the correct free theory limit}

For the $0$-form in adjoint representation of $ho(1|2:[3,2])$, the equations of motion and the gauge transformation are
\begin{equation}\label{x34c}
	d W^{\alpha}  =\frac{1}{2}\bar{f}^{\alpha}_{\beta\gamma}W^{\beta} \wedge W^{\gamma},\;\;\;\;\;\;\;d \Phi^{\alpha}=\bar{f}^{\alpha}_{\beta\gamma}W^{\beta} \Phi^{\gamma}
\end{equation}
and 
\begin{equation}\label{x34cg}
	\delta_{\epsilon} W^{\alpha}=d	\epsilon^{\alpha}+ \bar{f}^{\alpha}_{\beta\gamma} \epsilon^{\beta} W^{\gamma},\;\;\;\;\;\;\;\delta_{\epsilon} \Phi^{\alpha}= \bar{f}^{\alpha}_{\beta\gamma}  \epsilon^{\beta}\Phi^{\gamma}
\end{equation}
with
\begin{equation}\label{12wsfor}
	\bar{f}^{\alpha}_{\beta[\gamma}\bar{f}^{\beta}_{\rho\sigma]}-\Phi^{\nu}\bar{f}^{\beta}_{\nu[\gamma}\frac{\partial \bar{f}^{\alpha}_{\rho\sigma]}}{\partial \Phi^{\beta}}=0.
\end{equation}
Expanding in terms of $\Phi^{\alpha}$,
\begin{equation}\label{fxgf}
	\bar{f}^{\alpha}_{\beta\gamma} = f^{\alpha}_{\beta\gamma}+f^{\alpha}_{\beta\gamma|\sigma_{1}}\Phi^{\sigma_{1}}+f^{\alpha}_{\beta\gamma|\sigma_{1}\sigma_{2}}\Phi^{\sigma_{1}}\Phi^{\sigma_{2}}+\cdots
\end{equation}
If we assume
\begin{equation}
	t_{\alpha}f^{\alpha}_{\beta\gamma|\sigma_{1}\cdots \sigma_{n}} = f(t_{\beta}, t_{\gamma}; t_{\sigma_{1}}\cdots t_{\sigma_{n}})~,~
\end{equation}
where $f(t_{\beta}, t_{\gamma}; t_{\sigma_{1}}\cdots t_{\sigma_{n}})$ is the sum of the product of $t_{\beta}, t_{\gamma}, t_{\sigma_{1}},\cdots, t_{\sigma_{n}}$ with $t_{\beta}$ and $t_{\gamma}$ antisymmetric, $t_{\sigma_{1}}\cdots t_{\sigma_{n}}$ symmetric, then 
\begin{equation}
t_{\alpha}	\bar{f}^{\alpha}_{\beta\gamma} =  f(t_{\beta},t_{\gamma})+ f(t_{\beta}, t_{\gamma}; \Phi)+f(t_{\beta}, t_{\gamma}; \Phi,\Phi)+\cdots 
\end{equation}
$\Phi=\Phi^{\alpha}t_{\alpha}$. With (\ref{fxgf}) plugged in (\ref{12wsfor}), comparing the coefficients order by order, the solution can only be
\begin{equation}\label{fgxso}
t_{\alpha}	\bar{f}^{\alpha}_{\beta\gamma} = [t_{\beta},t_{\gamma}]F(\Phi), \;\;\;\;\;\;\;\; \textnormal{or}\;\;\;\;\;\;\;\;t_{\alpha}	\bar{f}^{\alpha}_{\beta\gamma} =F(\Phi) [t_{\beta},t_{\gamma}], 
\end{equation}
where $F(\Phi)$ is an arbitrary function of $\Phi$ with $F(0)=1$. Plug (\ref{fgxso}) into (\ref{12wsfor}), we can see (\ref{12wsfor}) is indeed satisfied. (\ref{x34c}) and (\ref{x34cg}) become 
\begin{equation}\label{fgxfo}
	d W  =[W,W]F(\Phi),\;\;\;\;\;\;\;d \Phi=[W,\Phi]F(\Phi)
\end{equation}
and 
\begin{equation}
	\delta_{\epsilon} W=d	\epsilon+ [\epsilon,W]F(\Phi),\;\;\;\;\;\;\;\delta_{\epsilon} \Phi=[\epsilon,\Phi]F(\Phi) ~,
\end{equation}
or
\begin{equation}
	d W  =F(\Phi)[W,W],\;\;\;\;\;\;\;d \Phi=F(\Phi)[W,\Phi]
\end{equation}
and 
\begin{equation}\label{fgxfo1}
	\delta_{\epsilon} W=d	\epsilon+F(\Phi) [\epsilon,W],\;\;\;\;\;\;\;\delta_{\epsilon} \Phi=F(\Phi)[\epsilon,\Phi] .
\end{equation}
\begin{equation}\label{fxgso}
	\bar{f}^{\alpha}_{\beta\gamma}=\left\langle [t_{\beta},t_{\gamma}]F(\Phi)|t^{\alpha}\right\rangle=f^{\sigma}_{\beta\gamma}\left\langle t_{\sigma}F(\Phi)|t^{\alpha}\right\rangle \;\;\;\;\;\textnormal{or}\;\;\;\;\;  \bar{f}^{\alpha}_{\beta\gamma}=\left\langle F(\Phi)[t_{\beta},t_{\gamma}]|t^{\alpha}\right\rangle=f^{\sigma}_{\beta\gamma}\left\langle F(\Phi)t_{\sigma}|t^{\alpha}\right\rangle. 
\end{equation}
Each $F(\Phi)$ gives a consistent deformation of $\bar{f}^{\alpha}_{\beta\gamma}=f^{\alpha}_{\beta\gamma}$. With the field redefinition $\Phi'=f(\Phi)$, (\ref{fgxfo})-(\ref{fgxfo1}) become
\begin{equation}
	d W  =[W,W]F[f^{-1}(\Phi')],\;\;\;\;\;\;\;d \Phi'=[W,\Phi']F[f^{-1}(\Phi')]
\end{equation}
and 
\begin{equation}
	\delta_{\epsilon} W=d	\epsilon+ [\epsilon,W]F[f^{-1}(\Phi')],\;\;\;\;\;\;\;\delta_{\epsilon} \Phi'=[\epsilon,\Phi']F[f^{-1}(\Phi')] ~,
\end{equation}
or
\begin{equation}
	d W  =F[f^{-1}(\Phi')][W,W],\;\;\;\;\;\;\;d \Phi'=F[f^{-1}(\Phi')][W,\Phi']
\end{equation}
and 
\begin{equation}
	\delta_{\epsilon} W=d	\epsilon+F[f^{-1}(\Phi')] [\epsilon,W],\;\;\;\;\;\;\;\delta_{\epsilon} \Phi'=F[f^{-1}(\Phi')][\epsilon,\Phi'] .
\end{equation}
Especially, when $f=F$, $F[f^{-1}(\Phi')] = \Phi'$. All of the consistent deformations are related to $F(\Phi)=\Phi$ by a field redefinition.

Until now, we have not made any assumption on the algebra $\{t_{\alpha}\}$, so (\ref{fxgso}) holds for the arbitrary algebra which is also a ring. Consider the $4d$ bosonic higher spin theory with the spin $s=0,1,2,\cdots$ and the algebra $g$, for $t_{\alpha} \in g$, $t_{\alpha}\sim t_{a_{1}\cdots a_{s-1},b_{1}\cdots b_{t}0\cdots 0} \sim y^{m}\bar{y}^{n}$ with $m+n=2(s-1)$, $t=|m-n|/2$. $\forall \; t_{\alpha}, t_{\beta} \in g$, $t_{\alpha}t_{\beta} \in g$, so (\ref{fgxfo})-(\ref{fgxfo1}) are well defined, but the truncation to the minimal bosonic higher spin theory is not possible. The theory does not have the correct free theory limit since (\ref{4587}) is not satisfied. Also, $R^{\alpha}_{(ab)\gamma}\neq 0$, the local Lorentz transformation is deformed.

\bibliographystyle{plain}

\end{document}